\documentclass[trackchanges,twocolumn,tighten,resetfootnote]{aastex7}

\newcommand{\ltsima}{$\; \buildrel < \over \sim \;$}
\newcommand{\simlt}{\lower.5ex\hbox{\ltsima}}

\newcommand{\hii}{H~{\sc ii}}
\def\arcdeg{\hbox{$^\circ$}}
\def\arcmin{\hbox{$^\prime$}}
\def\arcsec{\hbox{$^{\prime\prime}$}}

\newcommand{\IRfluxratio}{$\langle\nu F_\nu\rangle_{24}^{ns}/\langle\nu F\nu\rangle_{70}$}

\usepackage{amsmath}
\usepackage{graphicx}

\shorttitle{Taming the Tarantula}
\shortauthors{Rodriguez et al.}

\graphicspath{{./}{figures/}}

\begin{document}

\title{Taming the Tarantula: How Stellar Wind Feedback Shapes Gas and Dust in 30 Doradus}

\correspondingauthor{Jennifer A. Rodriguez}
\email{rodriguez.1392@osu.edu}

\author[0000-0003-1560-001X]{Jennifer A. Rodriguez} 
\affil{Department of Astronomy, The Ohio State University, 140 W. 18th Ave., Columbus, OH 43210, USA}
\affil{Center for Cosmology and AstroParticle Physics, The Ohio State University, 191 W. Woodruff Ave., Columbus, OH 43210, USA}
\email{rodriguez.1392@osu.edu}

\author[0000-0002-1790-3148]{Laura A. Lopez}
\affil{Department of Astronomy, The Ohio State University, 140 W. 18th Ave., Columbus, OH 43210, USA}
\affil{Center for Cosmology and AstroParticle Physics, The Ohio State University, 191 W. Woodruff Ave., Columbus, OH 43210, USA}
\email{lopez.513@osu.edu}

\author[0000-0002-0041-4356]{Lachlan Lancaster}
\thanks{Simons Fellow}
\affiliation{Department of Astronomy, Columbia University,  550 W 120th St, New York, NY 10025, USA}
\affiliation{Center for Computational Astrophysics, Flatiron Institute, 162 5th Avenue, New York, NY 10010, USA}
\email{llancaster@flatironinstitute.org}

\author[0000-0003-4423-0660]{Anna L. Rosen}
\affil{Department of Astronomy, San Diego State University, 5500 Campanile Dr, San Diego, CA 92182, USA}
\affil{Computational Science Research Center, San Diego State University, 5500 Campanile Dr, San Diego, CA 92182, USA}
\email{alrosen@sdsu.edu}

\author[0000-0001-6576-6339]{Omnarayani Nayak}
\affil{Space Telescope Science Institute, 3700 San Martin Drive, Baltimore, MD 21218, USA}
\email{omnarayani.nayak@nasa.gov}

\author[0000-0002-2644-0077]{Sebastian Lopez}
\affil{Department of Astronomy, The Ohio State University, 140 W. 18th Ave., Columbus, OH 43210, USA}
\affil{Center for Cosmology and AstroParticle Physics, The Ohio State University, 191 W. Woodruff Ave., Columbus, OH 43210, USA}
\email{lopez.764@osu.edu}

\author[0000-0002-7643-0504]{Tyler Holland-Ashford}
\affil{Astrophysics Division, NASA Goddard Space Flight Center, Greenbelt, MD 20771, USA}
\email{tyler.e.holland-ashford@nasa.gov}

\author[0009-0005-7228-9290]{Trinity L. Webb}
\affil{Department of Astronomy, The Ohio State University, 140 W. 18th Ave., Columbus, OH 43210, USA}
\affil{Center for Cosmology and AstroParticle Physics, The Ohio State University, 191 W. Woodruff Ave., Columbus, OH 43210, USA}
\email{webb.916@osu.edu}

\begin{abstract}

Observations of massive star-forming regions show that classical stellar wind models over-predict the luminosity of the X-ray emitting gas, indicating a significant fraction of wind energy is lost. In this paper, we present a multi-wavelength analysis of the giant \hii\ region 30 Doradus and its central star cluster R136 using 2~Ms of Chandra X-ray Observatory data, combined with James Webb Space Telescope and Hubble Space Telescope imaging and Spitzer spectral-energy distributions, to investigate how the hot gas energy is lost through turbulent mixing, radiative cooling, and physical leakage. We compare the spatial and spectral properties of the hot gas with those of the warm ionized gas and dust. We find no significant correlation between the dust and hot gas temperatures, suggesting they are not directly coupled and that the dust resides in the swept-up shells where it is heated radiatively. H$\alpha$ and X-ray surface brightness profiles show that the X-rays peak interior to the H$\alpha$ shells, demonstrating partial confinement of the hot gas. The fragmented shell structure and bright X-ray interior that declines near the H$\alpha$ shell reflect efficient cooling from turbulent mixing at the hot-cold interface. We compare against recent simulations of stellar-feedback driven bubbles which have broad agreement with the morphology of the X-ray and H$\alpha$ emission, but the simulations produce a dip in the interior X-ray surface brightness and a lack of hard X-rays compared to the observations. These differences may suggest thermal conduction is important as mass-loading of the hot bubble could reproduce the X-ray observables.

\end{abstract}


\section{Introduction} \label{sec:intro}

Massive stars ($M \gtrsim 8~M_{\odot}$) form within clustered environments in molecular clouds and inject energy and momentum into the surrounding interstellar medium (ISM) through a variety of stellar feedback processes \citep{Krumholz19}. Among these feedback modes, massive stars contract quickly achieving high luminosities as they are actively forming, and launch fast, line-driven stellar winds \citep{Vink01}. These winds collide with the ISM, generating shock-heated $T_{\rm X}\sim10^7$~K material that rapidly expands, sweeping up the surrounding ISM and producing tenuous X-ray emitting cavities, commonly referred to as bubbles and superbubbles \citep{Bruhweiler80, Townsley06a}. 

Early X-ray observations from \textit{Einstein} and \textit{ROSAT} have detected diffuse X-ray emission from superbubbles \citep{Chu90,Wang91,Dunne01}, suggesting that stellar wind and supernova (SN) feedback shock-heats and fills these cavities with hot gas. With the addition of \textit{Chandra X-ray Observatory}, the diffuse X-ray emission has been detected and spatially resolved, separating point source and diffuse emission, across massive stars clusters (MSCs) and star-forming regions in the Magellanic Clouds \citep{Townsley06a,Lopez11,Lopez14,Webb24} and Milky Way \citep{Moffat02,Townsley03,Ezoe06,Townsley11,Townsley14,Townsley18}, revealing the complex morphology between the hot gas and surrounding cooler material.

While the integrated kinetic energy from stellar winds over the lifetimes of massive stars is similar to that produced by their resulting SNe \citep{Agertz13}, the dynamical impact of stellar wind feedback on the surrounding ISM and how it evolves is not well understood \citep{HC09, Lopez11, Rosen14}. There are several models that account for the X-ray luminosity in bubbles and superbubbles produced by stellar winds and SNe. The classical models of \cite{Castor75} and \cite{Weaver77} assume that the shock-heated gas is entirely confined within a dense, cool shell of swept-up ISM. Compared to the absorption-corrected X-ray luminosities from observations ($L_{\rm X}\sim10^{32}-10^{36}\ {\rm erg\ s^{-1}}$), these models produce high X-ray luminosities due to the effects of the thermal conduction between the hot interior and the cold shell. Alternatively, the model from \cite{CC85} describes a steady, free-flowing wind and ignores the surrounding ISM, predicting lower X-ray luminosities than observed. 

There are various channels by which stellar wind energy and X-ray luminosity are lost from \hii\ regions. The intermediate model from \cite{HC09} suggests that hot gas is partially confined and escapes as a free-flowing wind from a porous shell through physical leakage. Another form of wind energy loss is dust mixing and heating, in which turbulence occurring at the hot-cold interface mixes dust grains into the hot gas. Eventually, these dust grains embedded in the hot gas are destroyed through thermal radiation and dust sputtering, with smaller grains destroyed more quickly \citep{Rosen14}. Turbulent mixing at the hot-cold interface can efficiently cool the hot gas carried into the interface and ultimately lead to wind luminosity losses through radiative cooling at the intermediate temperatures created in the mixed gas \citep{Lancaster21a, Lancaster21b, Lancaster24}. 

These energy loss channels are traced by multi-wavelength analyses combining X-ray, optical, and infrared (IR) observations. H$\alpha$ observations trace the warm ($\sim10^4$~K) ionized gas at the interfaces where the hot gas interacts with the cooler ISM, highlighting the regions of shell fragmentation and leakage. IR observations reveal the dust heated by stellar radiation and highlight where dust mixes into the hot-cold interface and can possibly be sputtered and destroyed by the hot gas. The IR emission also reflects regions of stochastic heating of small dust grains, making it difficult to distinguish between radiative and collisional heating processes that can ultimately lead to grain destruction. Together, these observations provide a view of how hot gas interacts with the surrounding ISM, tracing leakage, cooling, and mixing that lead to potential stellar wind energy losses from \hii\ regions.

30 Doradus (hereafter, 30 Dor) is the most luminous and massive \hii\ region in the Local Group. It is powered by the central, young ($\sim$1--2 Myr) super-star cluster (SSC) R136 \citep{Hunter95,Crowther16,Bestenlehner20}, with a bolometric luminosity of $7.8 \times 10^7 L_{\sun}$ \citep{Malumuth94}. R136 hosts a rich stellar population of high-mass stars with initial stellar masses up to and exceeding 100~$M_\odot$ \citep{Crowther16, Bestenlehner20} and an initial stellar mass of $\geq$250~$M_\odot$ for the most massive star known, R136a1 \citep{Brands22}. 30 Dor was observed for 2-Ms as part of the Chandra X-ray Visionary Program ``The Tarantula--Revealed by X-rays" (T-ReX; \citealt{Townsley24}). T-ReX observed 30 Dor 51 times with Chandra ACIS from 2014 May 3 to 2016 January 22. The survey also included three archival Chandra ACIS-I observations totaling 92-ks from January 2006 \citep{Lopez11, Townsley14}. Previous X-ray studies of 30 Dor investigated the large-scale diffuse structure and massive stars using $\sim$24~ks of Chandra observations not included in the T-ReX survey \citep{Townsley06a, Townsley06b}. With 30 Dor's location in the Large Magellanic Cloud providing a low absorbing column ($\sim10^{21}$~cm$^{-2}$; \citealt{Boer80, Bestenlehner20, Brands22}) and nearly face-on orientation \citep{Marel01, Subramanian10, Choi18} at a distance of $\sim$50~kpc \citep{Grijs14}, we can get a detailed look at the morphology of the hot gas and comparison to the surrounding colder ISM to explore how the stellar wind energy is lost. 

In this paper, we investigate the stellar wind energy loss mechanisms described above from the young MSC at the center of 30 Dor. We present a multi-wavelength (X-ray, IR, and optical) analysis to probe the properties of the shock-heated wind gas and its relationship to the surrounding warm, photo-ionized gas and the dust contained within it. This paper is organized as follows. Section~\ref{sec:observations} presents the multi-wavelength observations and our methods to assess stellar wind feedback energy loss. In Section~\ref{sec:analysis&results}, we show our results of our multi-wavelength comparison of the hot gas, dust, and warm ionized gas, as well as comparison to simulations. In Section~\ref{sec:Discussion}, we discuss the implications of our results regarding dust mixing (Section~\ref{sec:Dust Mixing}), turbulent mixing (Section~\ref{sec:turbulent mixing}), and leakage (Section~\ref{sec:leakage}). We summarize our findings in Section~\ref{sec:conclusions}.

\section{Observations and Data Reduction} \label{sec:observations}

\subsection{X-ray Data} \label{sec:X-ray Data}

We downloaded, reprocessed, and reduced the data using Chandra Interactive Analysis Observations ({\sc ciao}) v4.14 \citep{CIAO2006} and {\sc caldb} v4.10.4. The list of Chandra observations, obtained by the \textit{Chandra X-ray Observatory}, are contained in the Chandra Data Collection (CDC) under~\dataset[doi:10.25574/cdc.505]{https://doi.org/10.25574/cdc.505}. The 54 observations were merged using the \textit{merge\_obs} function. Point sources were detected using the \textit{wavdetect} command and removed from the image using \textit{dmfilth}. Our approach differs from the analysis of \cite{Townsley18,Townsley24} which used the ACISExtract (\textit{AE}; \citealt{Broos10,Broos12, Broos22}) software package to define point sources with polygon regions. Because polygon regions are not supported in the standard {\sc ciao} pipeline procedures for extracting and filling point sources, we used the \textit{wavdetect} point sources and filled them to create the exposure-corrected images. For the spectral analysis, we excluded the \textit{AE}-identified point sources. 

\begin{figure*}
    \centering
\includegraphics[width=0.38\paperwidth]{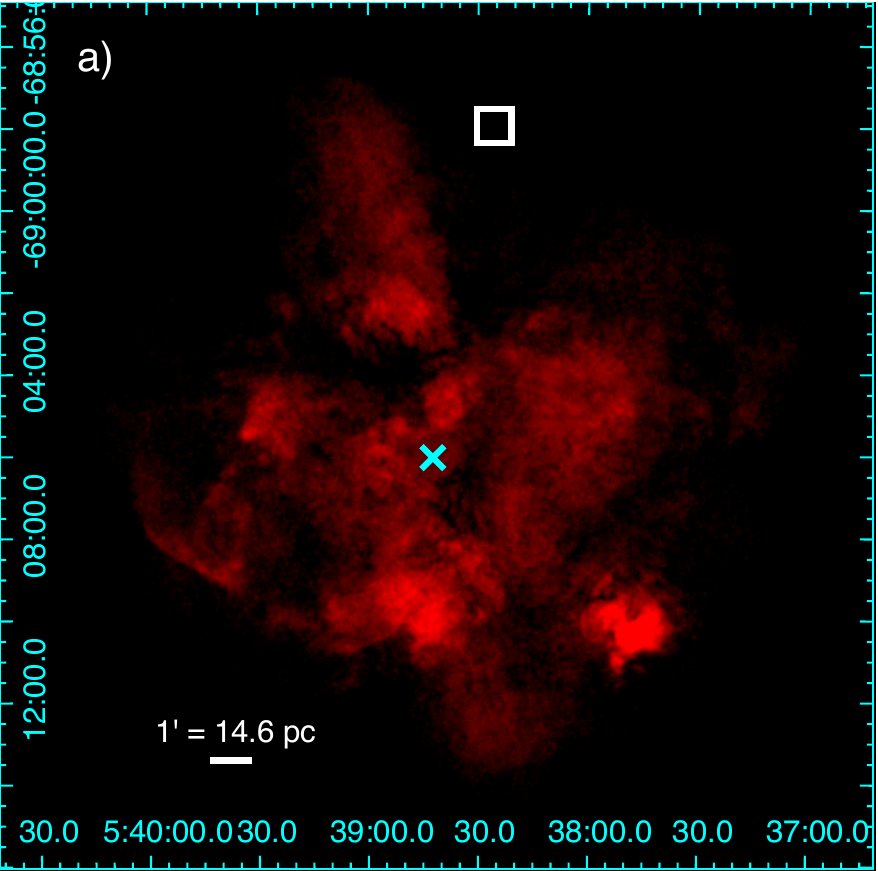}
\includegraphics[width=0.38\paperwidth]{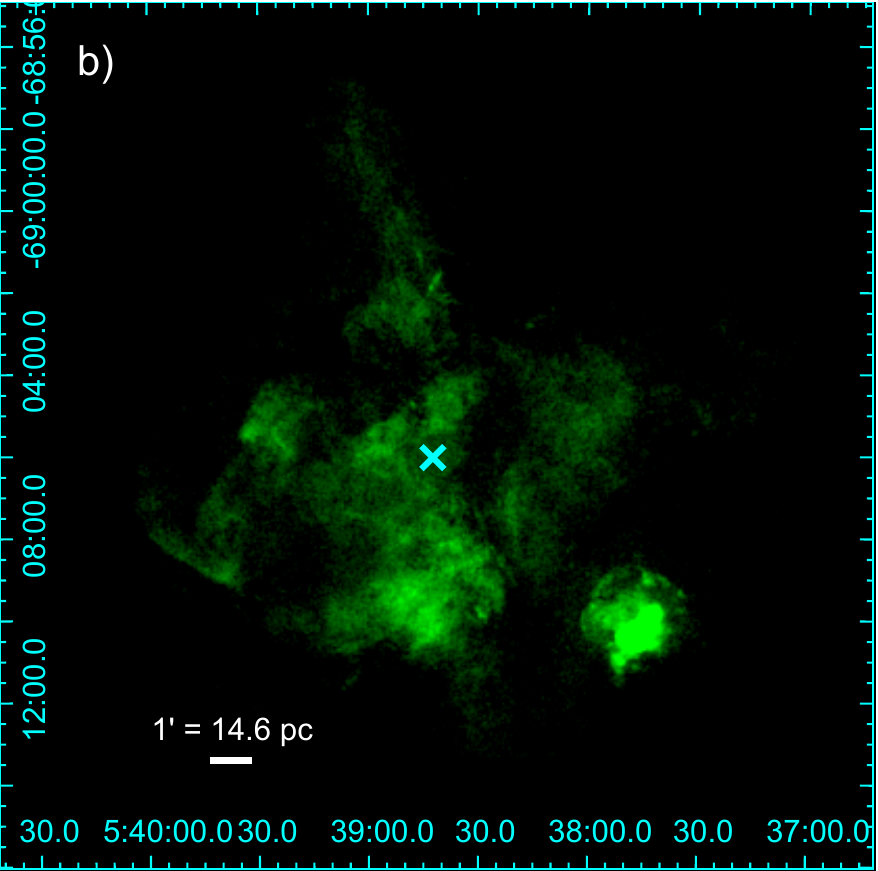}
\includegraphics[width=0.38\paperwidth]{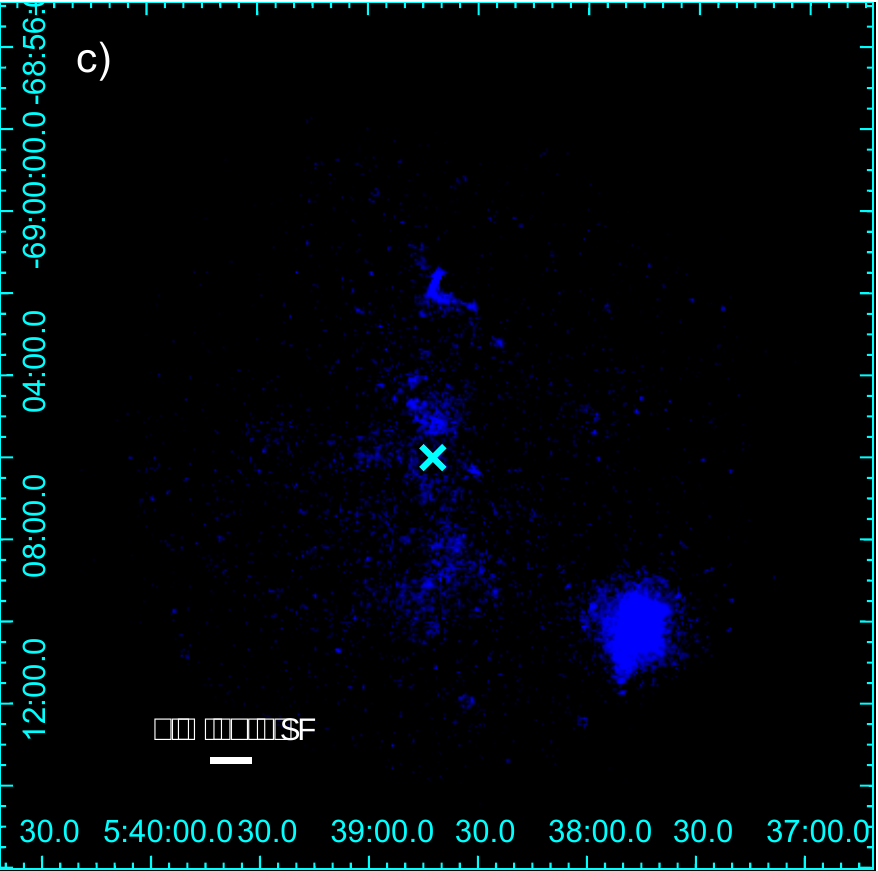}
\includegraphics[width=0.38\paperwidth]{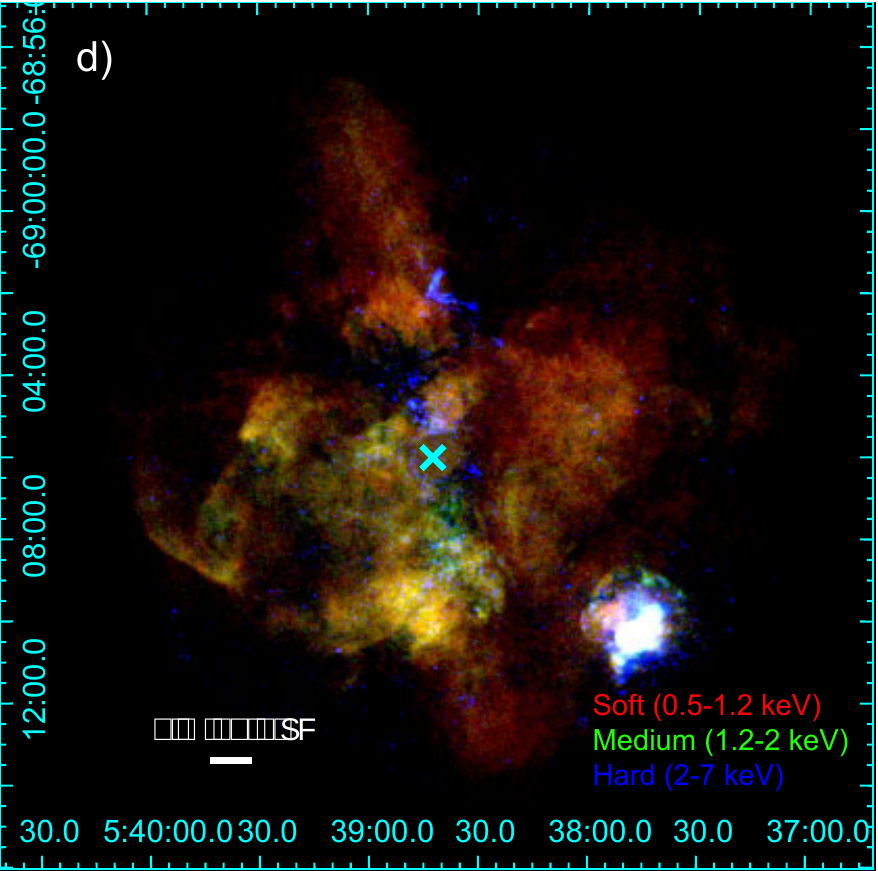}
    \caption{Composite X-ray images from 2 Ms of Chandra observations of 30 Doradus showing the complex structure of at least five superbubbles. a) Soft X-ray (0.5$-$1.2 keV) image of 30 Doradus. The white 50\arcsec$\times$50\arcsec\ box was used for background-subtraction. b) Medium X-ray (1.2$-$2 keV) image of 30 Doradus. c) Hard X-ray (2$-$7 keV) image of 30 Doradus. d) Three-color X-ray image of 30 Doradus, where soft X-rays (0.5$-$1.2 keV) are in red, medium X-rays (1.2$-$2 keV) are in green, and hard X-rays (2$-$7 keV) are in blue. The cyan X marks the location of R136, and the cluster is removed from the images. At a distance of 50~kpc, 1\arcmin\ is about 14.6~pc, as shown by the scale bar. The image is 22\arcmin$\times$22\arcmin. North is up, and East is left.}
    \label{fig:3color_xray}
\end{figure*}

Figure~\ref{fig:3color_xray} shows the exposure-corrected images for the soft (0.5$-$1.2 kev) X-rays in red, the medium (1.2$-$2 keV) X-rays in green, and the hard (2$-$7 keV) X-rays in blue. 30 Dor consists of at least five superbubbles \citep{Townsley06a} filled with softer X-ray emission. The complex structure of the bubbles was created by past and ongoing massive star formation. R136 is too young to have experienced SNe; however, 30 Dor has other stellar clusters of varied ages, including the $\sim$20~Myr star cluster Hodge 301 \citep{Grebel00} located $\sim$3\arcmin\ northwest of R136, where SNe have likely occurred and contributed to the diffuse X-ray emission. 30 Dor also has sources that produce hard X-ray emission, including the supernova remnant N157B, which contains the the pulsar PSR J0537-6910 \citep{Chen06, Chen23} and its associated pulsar wind nebula located $\sim$6\arcmin\ southwest of R136.

\subsection{X-ray Spectroscopy} \label{sec:X-ray Spectroscopy}

\begin{figure*}
    \centering
\includegraphics[width=0.4\paperwidth]{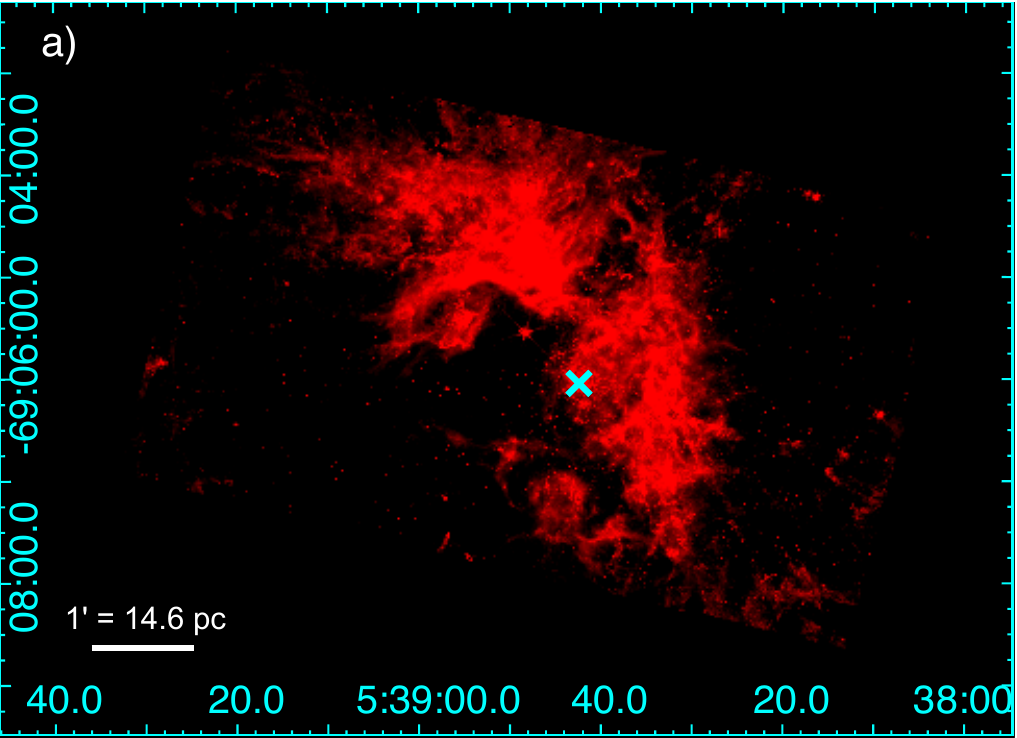}
\includegraphics[width=0.4\paperwidth]{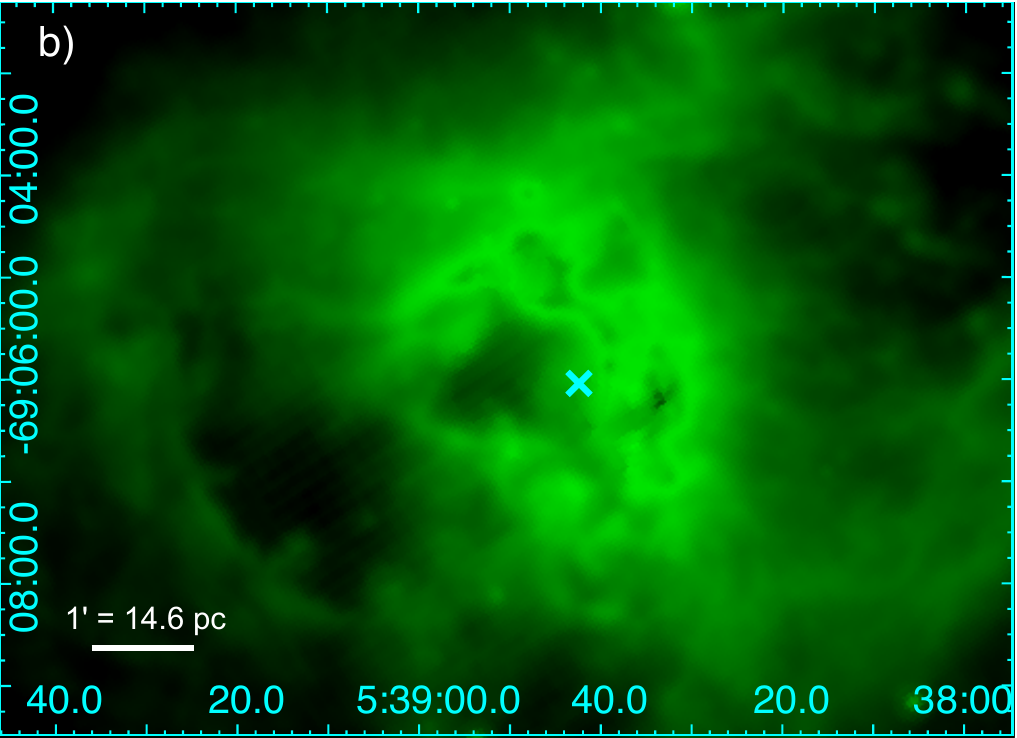}
\includegraphics[width=0.4\paperwidth]{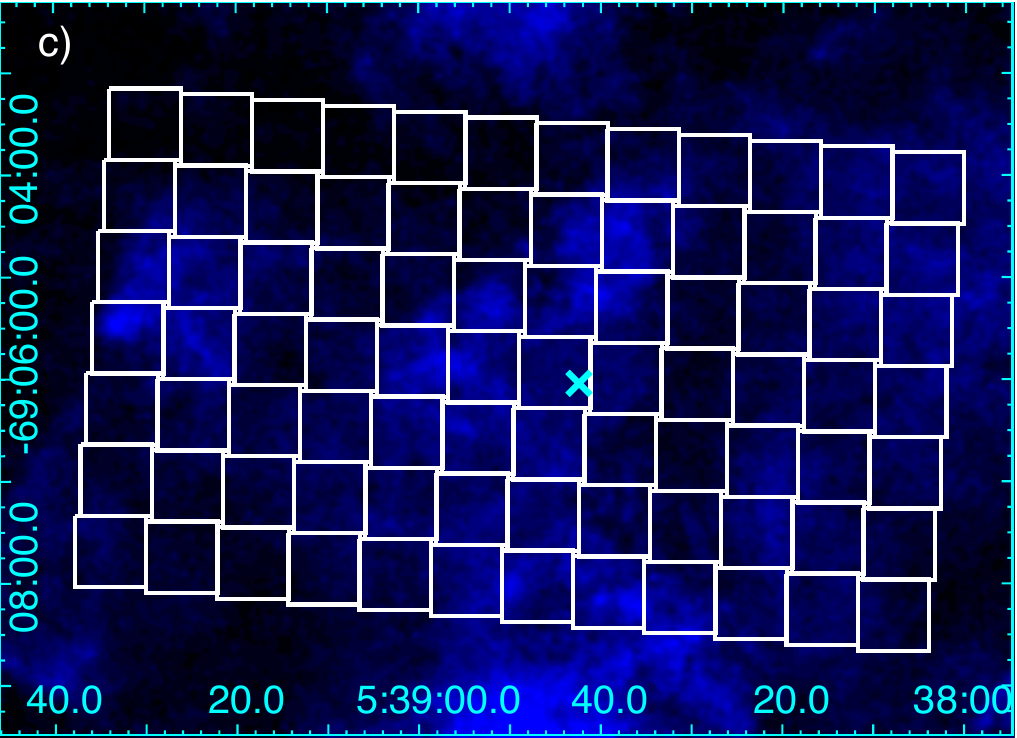}
\includegraphics[width=0.4\paperwidth]{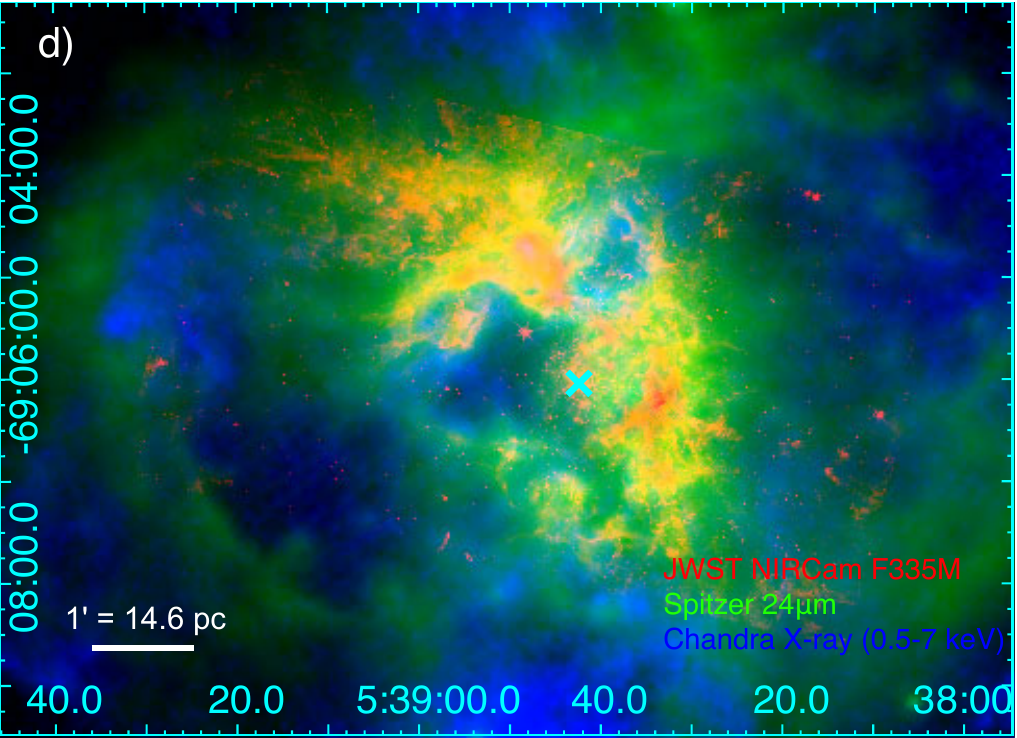}
    \caption{a) JWST NIRCam F335M observation of R136, tracing PAH emission. b) Spitzer 24$\mu$m observation, tracing warm dust heated by star formation. c) Chandra broad-band (0.5$-$7~keV) X-rays, tracing the hot gas, zoomed in on R136. The 84 (42\arcsec$\times$42\arcsec) regions used for X-ray spectral extraction are shown in white. d) 3-color image of 30 Doradus, with JWST NIRCam F335M in red, Spitzer MIPS 24$\mu$m in green, and broad-band X-rays (0.5$-$7~keV) in blue. The cyan X marks the location of R136. At a distance of 50~kpc, 1\arcmin\ is about 14.6~pc as shown by the scale bar. Images are 7\arcmin$\times$9\arcmin. North is up, and East is left. The hot gas fills the IR cavities, and the dust overlaps with the X-ray emission, suggesting that the dust may either occupy the same volume as the hot gas or remained confined to the shell.}
    \label{fig:84regions}
\end{figure*}

For our spectral analysis, we use 35 of the 54 T-ReX observations totaling $\sim$1.67~Ms of exposure time. We do not include additional observations with integration times below 30~ks each as their lower signal result in larger uncertainties in our spectral analysis. We extracted spectra using the {\sc ciao} command \textit{specextract} from 84 regions of area 42\arcsec$\times$42\arcsec\ (see Figure~\ref{fig:84regions}), with $\sim$1100--12000 net, full-band (0.5$-$7.0 keV) counts across the 35 observations. In this work, we adopt $1\arcsec\approx 0.24$~pc and $1\arcmin\approx 14.6$~pc at the distance of the LMC of 50~kpc\footnote{\cite{Pietrzynski19} constrained the distance to the LMC to 49.6$\pm$0.5~kpc.}. In order to constrain and compare the hot gas properties to those of the dust, we set the regions sizes to overlap with the spatially-resolved dust emission spectral energy distributions (SEDs) from \cite{Chastenet19} and with the JWST NIRCam field of view from the Early Release Observations (ERO) Program (ID 2729). The 3615 point sources identified with \textit{AE} in \cite{Townsley24} were excluded when extracting the spectra. The 84 regions were background-subtracted using a 50\arcsec$\times$50\arcsec\ region located on the same ACIS chip as the source regions $\sim$7\arcmin\ NW of R136 (see Figure~\ref{fig:3color_xray}).

The background-subtracted spectra from the 35 observations were jointly fit from 0.5$-$3~keV with XSPEC \citep{XSPEC} version 12.11.1. We focused on the soft X-rays to limit contribution from non-thermal sources. Our model includes an energy-independent constant ({\sc const}), two absorption components ({\sc phabs*vphabs}), and one optically thin plasma component ({\sc apec}). The constant was allowed to vary across all observations. The {\sc phabs} component was frozen to the column density of the Milky Way in the direction of 30 Dor, $N_{\rm H}^{\rm MW} = 6.44\times 10^{20}\ {\rm cm}^{-2}$ \citep{HI}. The {\sc vphabs} component, corresponding to the column density of the LMC toward 30 Dor, was allowed to vary. Solar abundances were adopted from \cite{Wilms00} and set to 0.5 solar metallicity, consistent with the LMC ISM \citep{Maggi16}. The temperature ($kT$) and the norm parameters from the {\sc apec} component were allowed to vary. 

In addition to the 84 42\arcsec$\times$42\arcsec regions, we extracted spectra from 60 regions to probe the hot gas properties in and around the H$\alpha$ cavities near R136 (see Figure~\ref{fig:hst_chandra}). Unlike the 84 regions we selected to compare to dust properties, the 60 regions were designed to trace the morphology of the H$\alpha$ shells and cavity interiors. The regions vary in size, with 43 regions of 25\arcsec$\times$25\arcsec, two regions of 50\arcsec$\times$50\arcsec, and 15 regions of 25\arcsec$\times$50\arcsec\ or 50\arcsec$\times$25\arcsec. The larger regions were adopted in certain locations to account for lower signal, which limited the hot gas temperature and column density constraints. Our regions have $\sim$1500$-$7700 net, full-band (0.5$-$7.0 keV) counts across the 35 observations.

\subsection{Infrared SED Modeling} \label{sec:Infrared SED}

We compare the hot gas properties derived from X-ray spectroscopic modeling to the dust properties from \cite{Lopez11} and \cite{Chastenet19} obtained using Spitzer IR SED modeling.

\cite{Lopez11} measured average IR flux densities for 441 regions of 35\arcsec$\times$35\arcsec\ across 30 Dor. Using Spitzer IRAC 3.6~$\mu$m and MIPS 24~$\mu$m and 70~$\mu$m images, they measured the average flux ratios (\IRfluxratio) from the warm dust emission with the stellar contribution removed and predicted the radiation field intensity ($U$) using the model from \cite{DL07}, where the radiation field is given by
\begin{equation}
    u_\nu = U u_\nu^{\rm ISRF},
    \label{eq:radiationfield}
\end{equation}
where $u_\nu$ is the energy density of the radiation field, $U$ is the dimensionless scale factor and $u_\nu^{\rm ISRF}$ is the interstellar radiation field for the local neighborhood, $8.65 \times 10^{-13}$~erg cm$^{-3}$ (from \citealt{Mathis83}).

\cite{Chastenet19} fitted SEDs across the LMC from the mid-IR through far-IR (3.6~$\mu$m $-$ 350~$\mu$m) using the combined observations from the Spitzer Space Telescope and Herschel Space Observatory. Using the dust emission model from \cite{DL07}, they created maps of the dust-mass-weighted average radiation field intensity ($\overline{U}$) and the fraction of the dust mass heated by the power-law distribution of the radiation field ($\gamma$) using a 42\arcsec\ ($\sim$10~pc) pixel size. Here, $\gamma$ represents the fraction of the dust in regions exposed to a stronger, variable radiation field, while the remaining fraction, $1-\gamma$, corresponds to the dust fraction exposed to the minimum, diffuse radiation field.

\subsection{H-alpha Imaging} \label{sec:H-alpha Imaging}

In order to explore the relationship of the hot gas with the warm diffuse ionized gas, we analyzed Hubble Space Telescope ACS/WFC F658N (corresponding to H$\alpha$) images of 30~Dor from the Hubble Tarantula Treasury Project \citep{Sabbi13, Sabbi16}. All the HST data used in this paper can be found in MAST: \dataset[10.17909/gyj0-3629]{http://dx.doi.org/10.17909/gyj0-3629}. Point sources and background regions were identified, in which we selected a scaling factor of 0.5 for background regions to control their sizes relative to the source regions. We removed the point sources from the H$\alpha$ image by replacing pixel values with those selected from the Poisson distribution of the background.

\begin{figure*}
    \centering
\includegraphics[width=0.27\paperwidth]{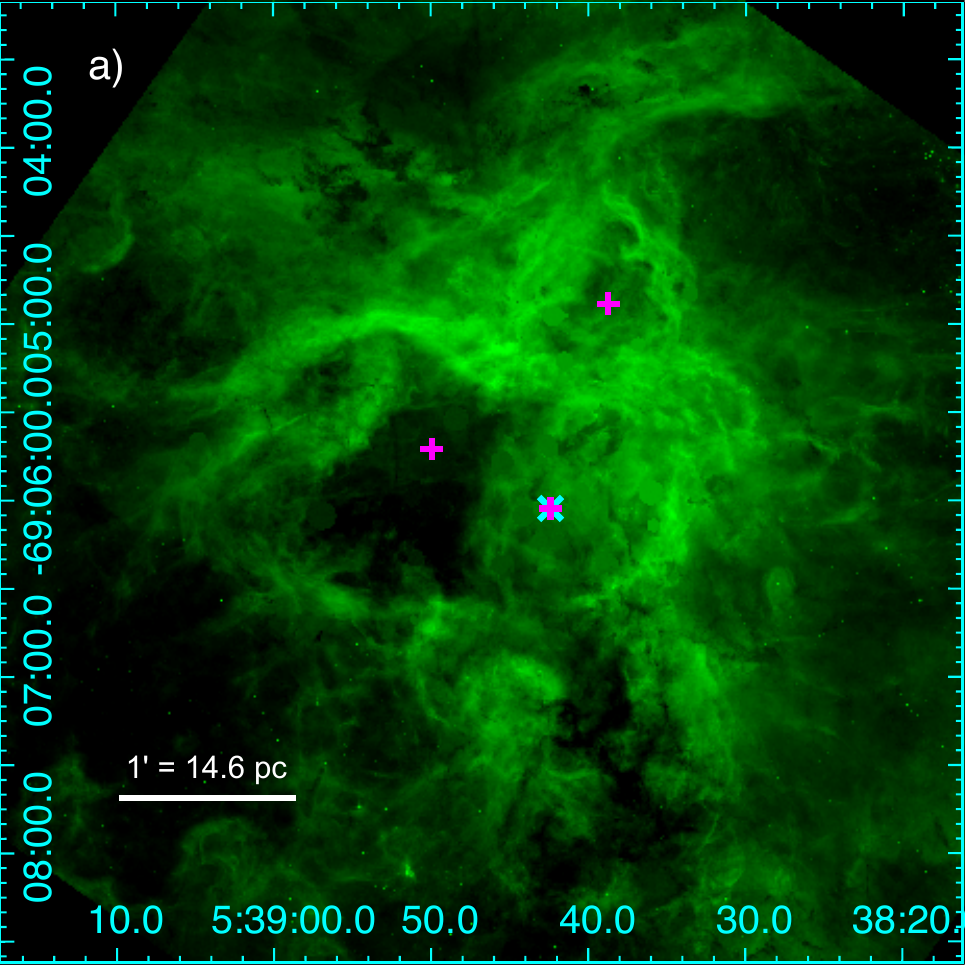}
\includegraphics[width=0.27\paperwidth]{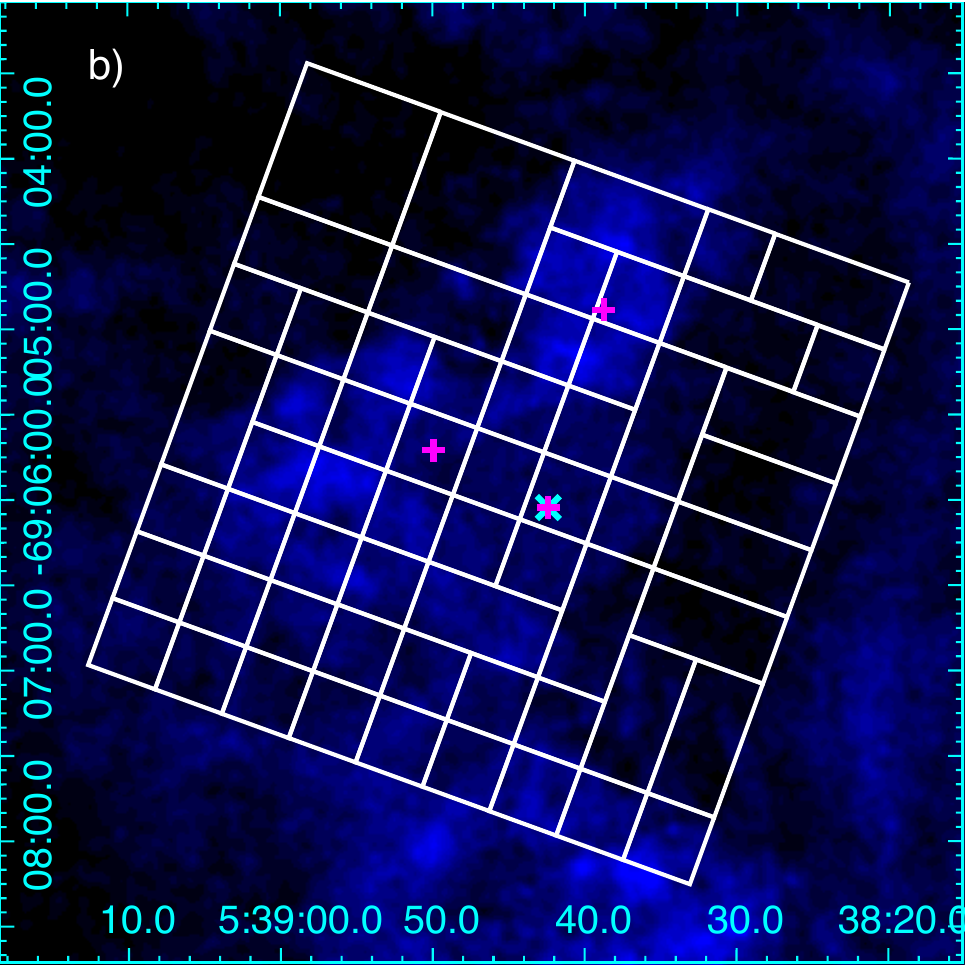}
\includegraphics[width=0.27\paperwidth]{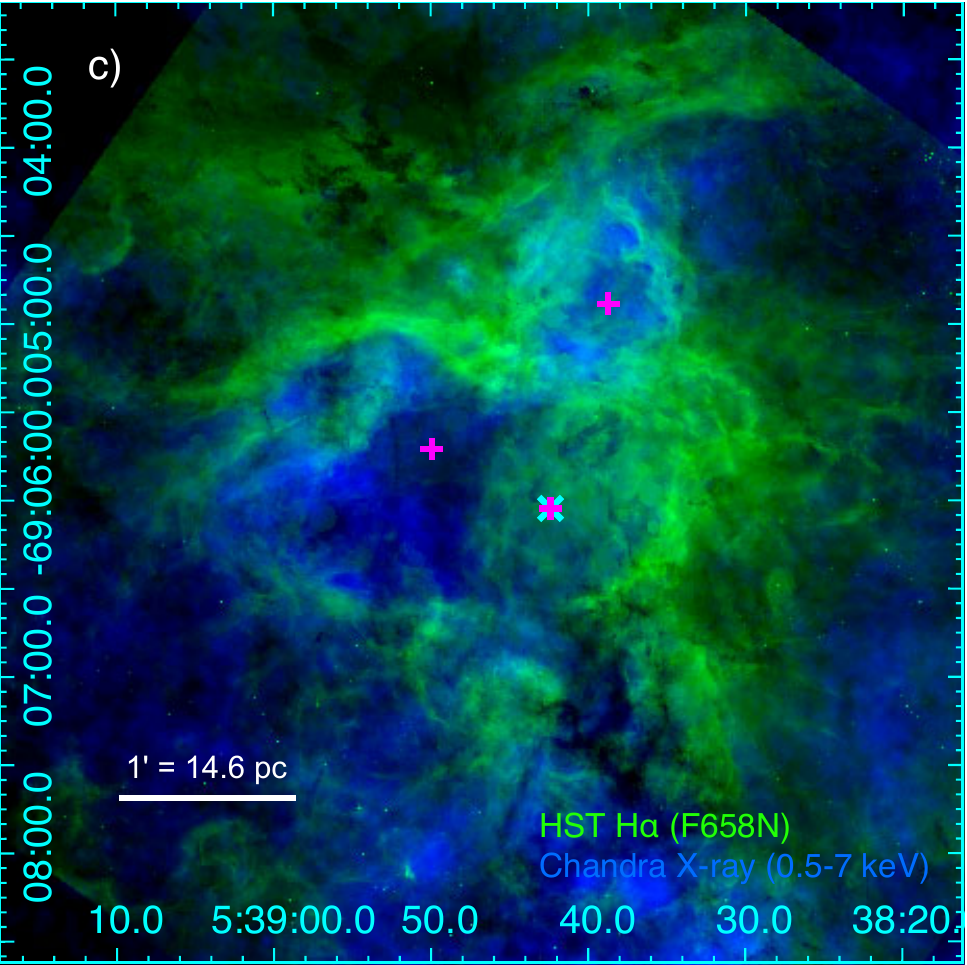}
    \caption{a) HST F658N image, corresponding to H$\alpha$. b) Chandra broad (0.5--7 keV) X-ray image with the regions used for the X-ray spectral extraction for the cavities seen in H$\alpha$ around R136. c) Composite image of the HST H$\alpha$ in green and Chandra broad X-ray in blue. The location of R136 is denoted by the cyan X and the magenta crosses mark the locations where we performed our surface brightness analysis (see Figures~\ref{fig:SB_R136}--\ref{fig:SB_smallcavity}). At a distance of 50~kpc, 1\arcmin\ is about 14.6~pc as shown by the scale bar. Images are 5.5\arcmin$\times$5.5\arcmin. North is up, and East is left. The hot gas fills the H$\alpha$ cavities, suggesting the hot gas may be confined by a shell of cooler gas.}
    \label{fig:hst_chandra}
\end{figure*}

\section{Analysis and Results} \label{sec:analysis&results}

\subsection{Image Analysis} \label{sec:imageanalysis}

Figure~\ref{fig:84regions} shows the three-color image zoomed in around R136 with JWST NIRCam F335M in red, Spitzer MIPS 24~$\mu$m in green, and broad (0.5$-$7~keV) X-rays (hot gas) in blue. JWST NIRCam F335M traces polycyclic aromatic hydrocarbon (PAH) emission, and Spitzer MIPS 24~$\mu$m traces emission of warm dust heated by star formation. Hot gas fills in the cavities seen in the IR observations. This is evident in the regions east and north of R136, where the diffuse X-ray emission occupies the cavities in the dusty ISM.

Similar to the morphology between the hot gas and dust, Figure~\ref{fig:hst_chandra} shows the complex nature between the HST H$\alpha$ (warm ionized gas) in green and the broad (0.5$-$7~keV) X-rays (hot gas) in blue. The diffuse X-ray emission also fills the warm gas cavities to the east and north of R136. The spatial distribution of the hot and warm gas suggests that the hot gas may be enclosed by a shell of cooler gas.

Figures~\ref{fig:84regions} and~\ref{fig:hst_chandra} indicate that the hot gas also fills the cavities within the cooler, dusty ISM. The diffuse X-ray emission appears bright or faint depending on whether it is located in front of or behind the surrounding ISM material. This suggests that the hot gas, resulting from stellar wind feedback from R136, creates bubbles enclosed by shells of dust and warm photoionized gas.

\subsection{X-ray Spectral Results} \label{sec:spectralresults}

\begin{figure}
    \centering
    \includegraphics[width=\columnwidth]{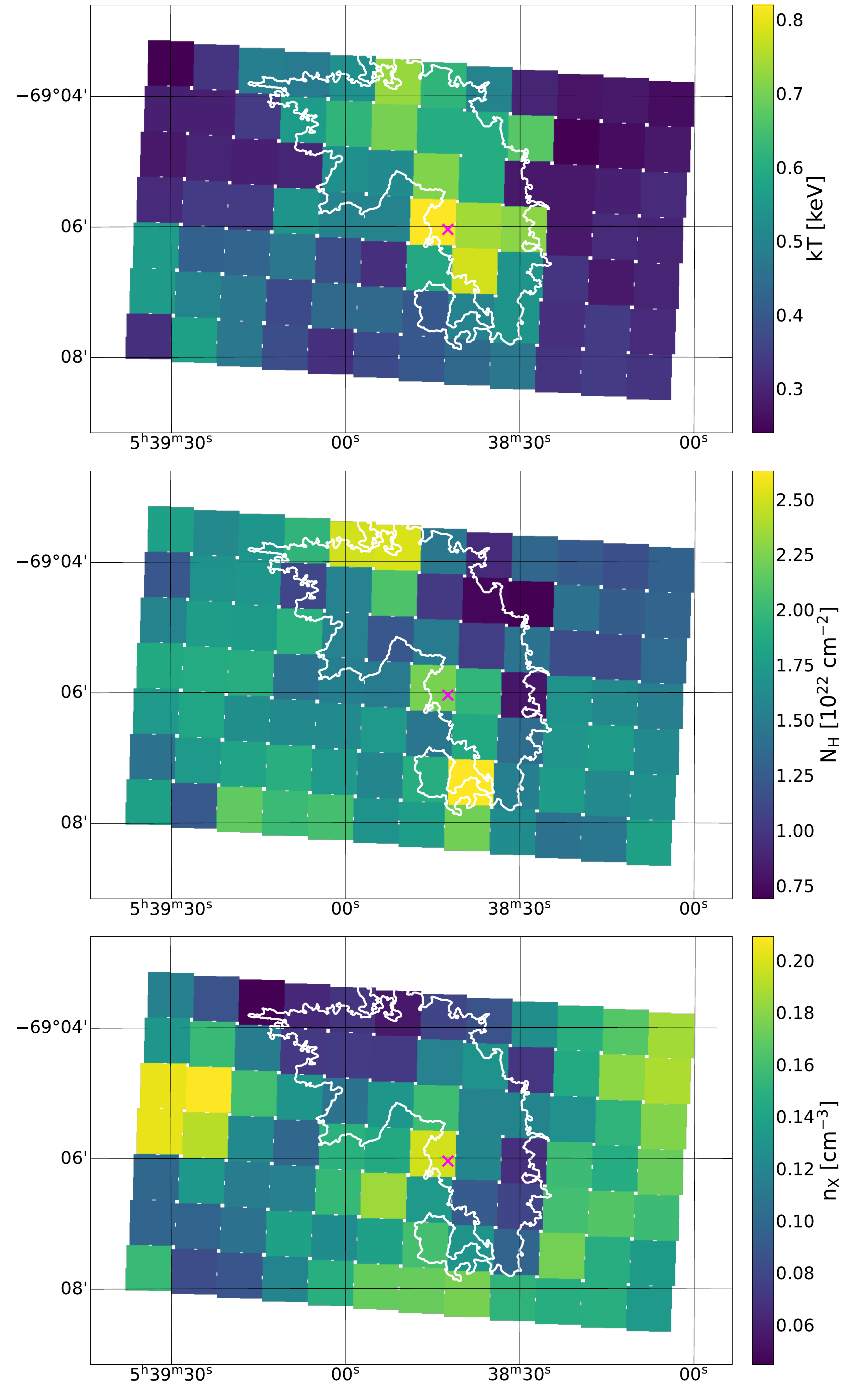}
    \caption{\textit{Top:} Map of the X-ray temperature $kT$ (keV) of 30 Dor derived from the spectral fits of the 84 regions (see Figure~\ref{fig:84regions}) selected to explore the relationship between the hot gas and dust. Typical errors of hot gas temperature were $\sim$10\%. \textit{Middle:} Map of the column density $N_{\rm H}$ ($\times10^{22}$ cm$^{-2}$) of 30 Dor, with $\sim$10\% error. \textit{Bottom:} Map of the X-ray electron density $n_{\rm X}$ (cm$^{-3}$) of 30 Dor. The white contour shows the JWST NIRCam F335M observation. Each region is 42\arcsec $\times$42\arcsec. The magenta X marks the location of R136. North is up, and East is left. The X-ray temperature peaks at R136, and higher X-ray temperatures coincide with the JWST NIRCam F335M contours, suggesting that the hot gas from stellar feedback may influence the dust through heating or destruction. }
    \label{fig:maps}
\end{figure}

Using the spectral extraction methods described in Section~\ref{sec:X-ray Spectroscopy}, we find the best-fit values for the X-ray temperature ($kT$) and intrinsic column density ($N_{\rm H}$). In Figure~\ref{fig:maps}, we plot the best-fit values for the 84 regions. Using the normalization from the {\sc apec} component, we calculate the number density ($n_{\rm X}$). The norm is defined as ${\rm norm} = 10^{-14}{\rm EM}/(4\pi D^2)$, where the emission measure is ${\rm EM}=\int n_{\rm X}n_{\rm H}dV$. For a fully ionized plasma primarily composed of hydrogen and helium, the number density is $n_{\rm X}=n_{\rm H}+2n_{\rm He} \approx 1.2n_{\rm H}$ \citep{Sarazin86}. Using this relation, we obtain $n_{\rm X}=(1.5\times10^{15}{\rm norm}D^2/(fV))^{1/2}$, in which $V$ is the volume, $f$ is the filling factor, and $D$ is the distance to 30 Dor. We calculate the volume of each square region by multiplying the area of the regions (42\arcsec\ on each side) by the path length through a spherical bubble of radius $\sim100$~pc centered on R136, where the radius is determined by the scale where 90\% of the H$\alpha$ emission is enclosed \citep{Lopez14}. We also assume a filling factor of 1, i.e., that the X-rays occupy the full volume of each region (though the filling factor may be small in some cases; see Section~\ref{subsec:sim_comp} for mock observations of simulations from \citealt{Lancaster25b}). Smaller filling factors would increase $n_{\rm X}$, so our derived values can be interpreted as lower limits on the hot gas number density.

We find best-fit X-ray gas temperatures of $kT=0.24-0.82$~keV ($\sim 2.8\times10^6-9.5\times10^6$~K), with the peak at the location of R136 (see Figure~\ref{fig:maps}). Regions of higher $kT$ overlap with the JWST NIRCam F335M observation (white contour), suggesting the hot gas may influence the dust, either through heating or by contributing to its destruction. We find best-fit absorbing columns of $N_{\rm H}=(0.69-2.63) \times 10^{22}$~cm$^{-2}$, with higher column densities toward regions where the X-rays appear dimmer. We find number densities of $n_{\rm X} = 0.04-0.21$~cm$^{-3}$, with greater values at the location of R136 and east of R136, where \cite{Townsley24} defined an area of plasma confinement and possible mixing, causing bright diffuse X-ray emission.

\begin{figure}
    \centering
    \includegraphics[width=\columnwidth]{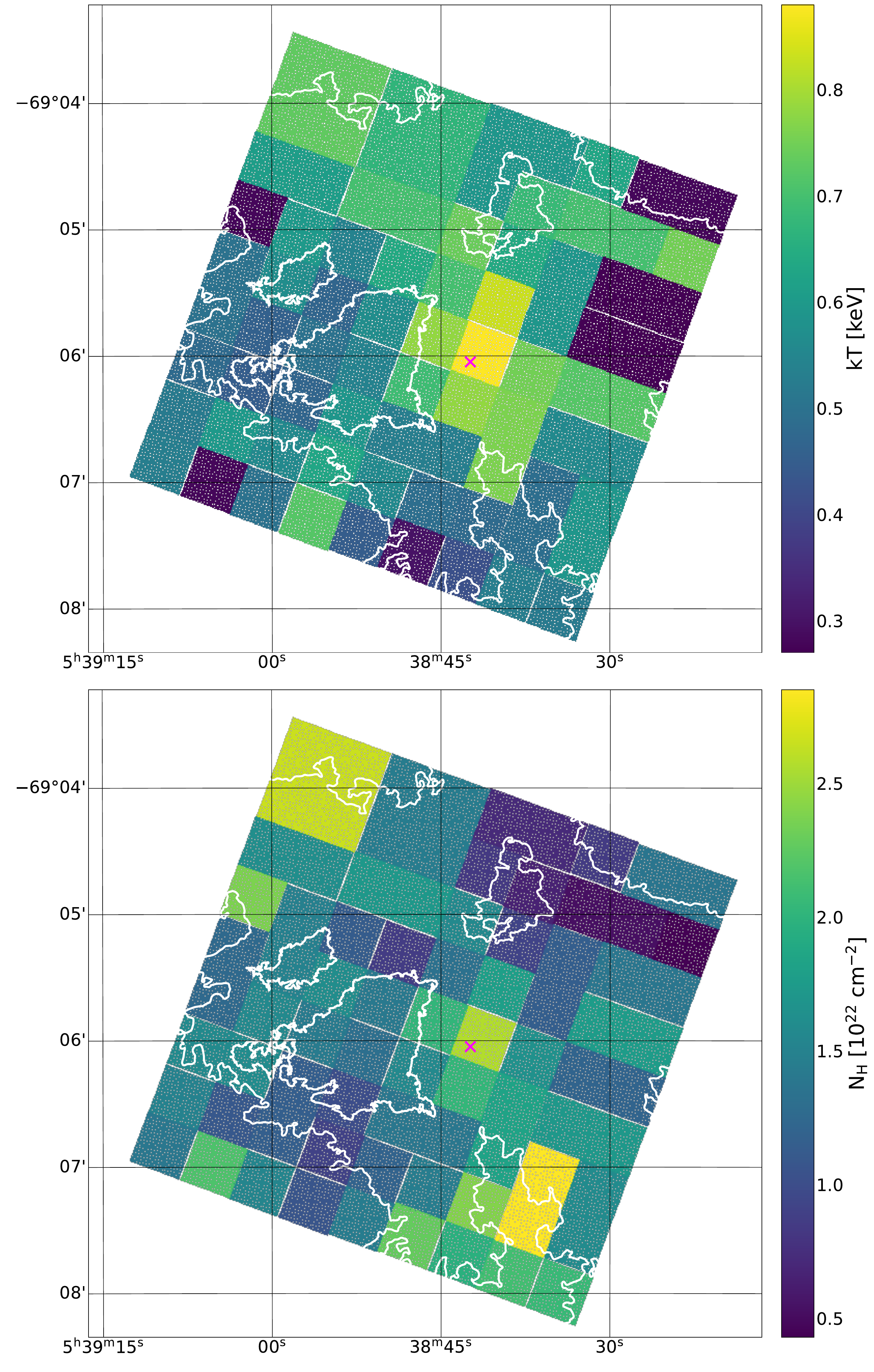}
    \caption{\textit{Top:} Map of the X-ray temperature $kT$ in keV from the spectral fits of the 60 regions surrounding R136 and nearby cavities to explore relationship between hot gas and warm gas (see Figure~\ref{fig:hst_chandra}), with $\sim$10\%--20\% error per region. \textit{Bottom:} Map of the column density $N_{\rm H}$ ($\times10^{22}$ cm$^{-2}$), with $\sim$10\% error per region. Region sizes vary and several are combined to account for low signal. The magenta X marks the location of R136. North is up, and East is left. HST H$\alpha$ contours are shown in white, highlighting the cavities east and north of R136. The values for each region correspond to the best-fit temperature from the 0.5$-$3 keV spectral analysis. The regions overlapping with the cavities east and north of R136 generally have low column densities and higher hot gas temperatures, though the eastern cavity shows relatively lower temperatures, suggesting a complex interplay of stellar feedback on the surrounding ISM. }
    \label{fig:map_smallregions}
\end{figure}

We also find the best-fit values for the X-ray temperature $kT$ and column density $N_{\rm H}$ for the 60 regions to compare to the H$\alpha$ cavities. We show the best-fit results in Figure~\ref{fig:map_smallregions}, with the white contour representing the distribution of the HST H$\alpha$. The X-ray temperatures and column densities vary across our regions: we find best-fit X-ray temperatures of $kT = 0.27-0.88$~keV ($\sim3.1\times10^6-1.0\times10^7$~K) and column densities of $N_{\rm H}=(0.43-2.85)\times10^{22}$~cm$^{-2}$.

Our results are broadly consistent with those of \cite{Townsley24}, who used tessellated extraction regions as well as a global region capturing the entirety of 30 Dor. They modeled the broad-band (0.5 $-$ 7.0 keV) X-ray spectra with multi-temperature fits to capture the soft and hard X-ray emission, while we adopt a single temperature model for the softer 0.5 $-$ 3 keV band. Despite the differences in these methods, the spectral X-ray results for their global diffuse X-ray spectrum are within our ranges, with \cite{Townsley24} reporting $kT = 0.28-0.94$~keV, $N_{\rm H}=(0.5-0.8)\times10^{22}$~cm$^{-2}$, and $n_{\rm X} = 0.03-0.07$~cm$^{-3}$.

\subsection{Comparison to Dust}

\begin{figure*}
    \centering
    \includegraphics[width=0.9\textwidth]{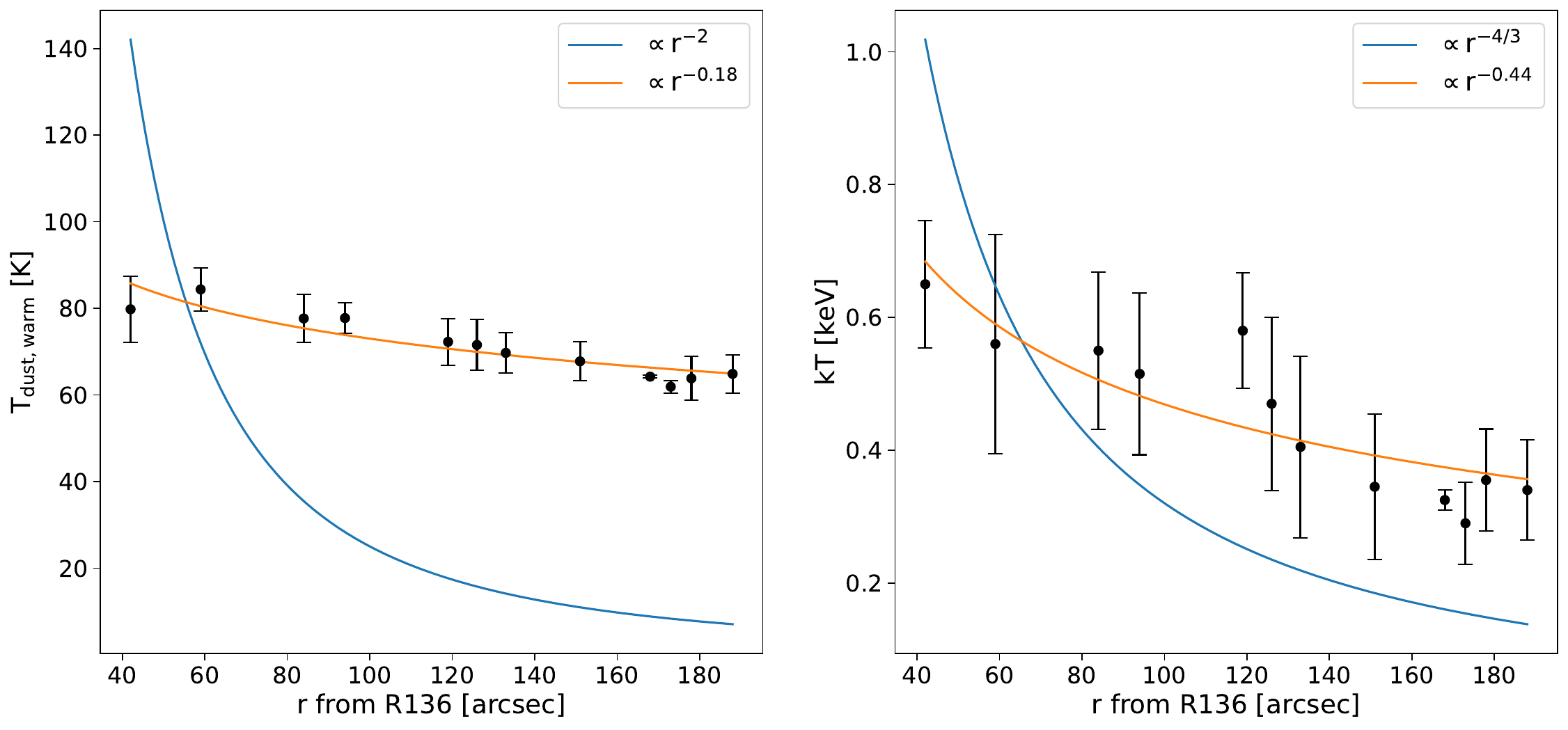}
    \caption{\textit{Left:} Warm dust temperature calculated from modeled radiation field intensity $U$ values from \cite{Lopez11} vs. distance $r$ from R136. The blue line demonstrates the $T_{\rm dust, warm} \propto r^{-2}$ relation for radiation, while the orange line represents the best-fit trend to the data, $T_{\rm dust, warm} \propto r^{-0.18}$. The error bars represent the 1$\sigma$ spread of the warm dust temperatures around the median values. \textit{Right:} Hot gas temperature from our spectral analysis in keV vs. distance from R136. The blue line represents adiabatic expansion from a wind-blown bubble $kT \propto r^{-4/3}$, while the orange line shows the best-fit trend to the data $kT \propto r^{-0.44}$. The error bars represent the spread of the hot gas temperatures around the median values. 50\arcsec=12~pc at the distance of the LMC. Both the warm gas and hot gas temperatures decline shallowly with distance from R136, with neither follows the expected theoretical scaling relation.}
    \label{fig:tempvsr}
\end{figure*}

We explore the relationship between the hot gas and the dust emission to investigate the loss of stellar wind energy due to dust mixing, heating, and destruction. In Figure~\ref{fig:tempvsr}, we compare the warm dust temperature (left) and hot gas temperature (right) and their dependence on distance from R136. We calculate the warm dust temperature from the radiation field intensity (see Equation~\ref{eq:radiationfield}) from \cite{Lopez11} using $T_{\rm dust, warm} = 20~U^{1/6}$~K. The regions used for spectral analysis in \cite{Lopez11} are 35\arcsec\ on a side and are smaller than our regions (42\arcsec\ on a side) described in Section~\ref{sec:X-ray Spectroscopy}. To account for this, we assign the average flux ratio to the regions with the most overlap. The distance to R136, $r$, is calculated using $r=\sqrt{(x_i-x_{R136})^2+(y_i-y_{R136})^2}$, where the centers of the regions in Figure~\ref{fig:84regions} are ($x_i$, $y_i$) and the center of the region containing R136\footnote{We mark the position of R136 as that of R136a1, adopting the coordinates from \cite{Kalari22} with R.A. 05:38:42.398 and decl. $-$69:06:02.86.} is ($x_{\rm R136}$, $y_{\rm R136}$).

We bin the dust and hot gas temperatures by distance from R136 and calculate the median and 1$\sigma$ spread of the temperatures within each bin. On the left panel of Figure~\ref{fig:tempvsr}, we plot the expected relation for warm dust temperature set by radiation from R136, $T_{\rm dust, warm} \propto r^{-2}$ (in blue) and the best-fit line to the warm dust temperatures, which is $T_{\rm dust, warm} \propto r^{-0.18}$ (in orange). On the right, we plot the expected relation for adiabatic expansion of the hot gas $kT \propto r^{-4/3}$ (in blue) and the best-fit line to the hot gas temperatures, which is $kT \propto r^{-0.44}$ (in orange). Overall, the warm dust and hot gas temperatures both decrease with distance from R136, with the warm dust temperature having a more gradual decline than the hot gas temperature. We discuss the discrepancy between the temperatures and simple models in Section~\ref{sec:Dust Mixing}.

\begin{figure*}[t]
    \centering    \includegraphics[width=\textwidth]{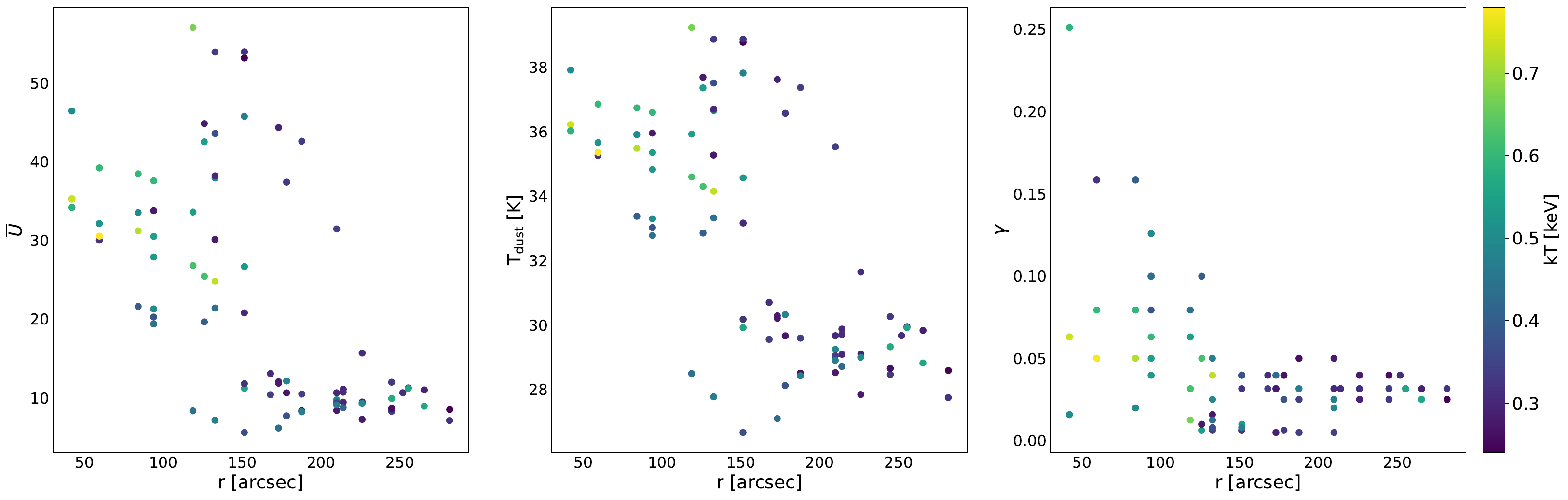}
    \caption{\textit{Left:} Dust-mass-weighted average radiation field intensity ($\overline{U}$) from \cite{Chastenet19} vs. distance $r$ from R136 in arcsec. \textit{Middle:} Dust temperature calculated from $\overline{U}$ vs distance from R136 $r$. \textit{Right:} Fraction of the dust mass heated by the power-law distribution of the radiation field from \cite{Chastenet19} vs distance form R136 $r$. The colorbar corresponds to the best-fit hot gas temperature $kT$ in keV. 50\arcsec$=$12~pc at the distance of the LMC. All three properties decrease out to $\sim$150\arcsec\ from R136 before leveling off, suggesting stronger radiation fields and dust heating closer to R136.}
    \label{fig:dust_properties}
\end{figure*}

We also compare the spatial trends of the hot gas temperature alongside the dust-mass-weighted average radiation field intensity ($\overline{U}$), the average dust temperature ($T_{\rm dust}$), and the fraction of the dust mass heated by the power-law distribution of the radiation field ($\gamma$) from \cite{Chastenet19}. We calculate the average dust temperature with $T_{\rm dust} = 20~\overline{U}^{1/6}$~K. The results are shown in Figure~\ref{fig:dust_properties}, with the dust parameters shown as a function of distance from R136, $r$. Each point is assigned a color based on the X-ray temperature, with hotter temperatures in yellow and cooler temperatures in dark blue, to provide additional context on the distribution of the hot gas compared to the dust. We omit the region containing R136 from this comparison as all three dust properties there are low due to the diffuse neutral gas along the line-of-sight which contaminates the measurements and modeling of these properties. 

Both $\overline{U}$ and $T_{\rm dust}$ decrease within $\sim$150\arcsec\ of R136, reflecting stronger radiation fields there, and decrease at larger radii ($r > 150\arcsec$) where the radiation fields are weaker. $\gamma$ also declines with distance from R136, reflecting the dust grains being exposed to weaker radiation fields out to  $r\sim$150\arcsec. The scatter in $\gamma$ indicates the local variations in the radiation field closer to R136. Hot gas temperatures are highest within 150\arcsec\ of R136 and decrease going outward, which is consistent with stellar wind feedback shock-heating the ISM in the regions surrounding R136, while the dust properties trace the radiation field.

\begin{figure}
    \centering
\includegraphics[width=\columnwidth]{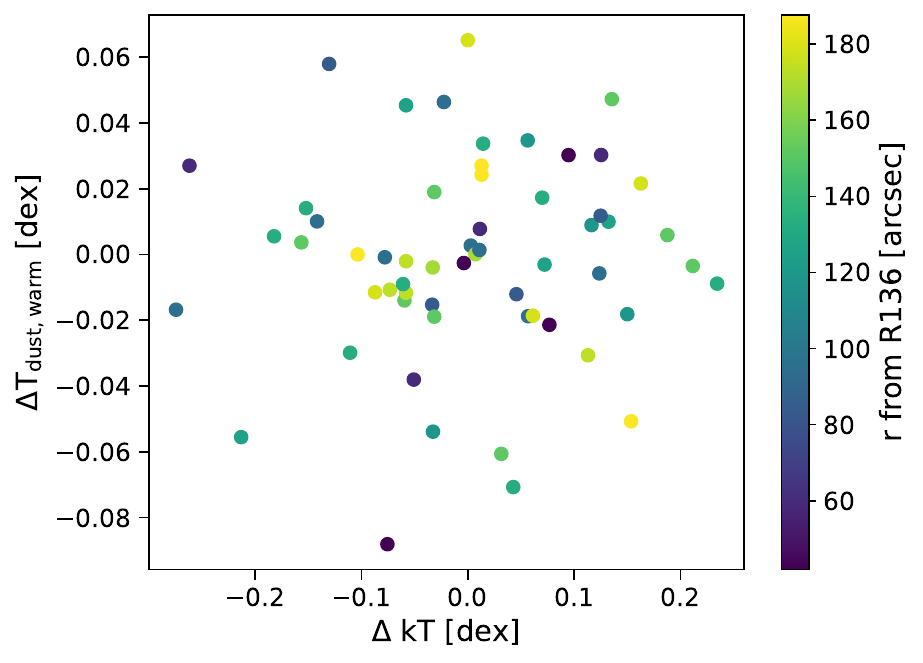}
    \caption{Distance-corrected residuals of warm dust temperature vs. hot gas temperature. The colorbar corresponds to distance from R136. No significant correlation is found between the temperatures, suggesting there is no causal relation between the hot gas and dust.}
    \label{fig:TdustvskT}
\end{figure}

In order to assess whether the correlation between X-ray and dust temperatures reflects a causal relationship, we account for their mutual correlation with distance from R136. We bin the temperatures by distance and calculate the median temperatures within each distance bin. We interpolate median temperatures and subtract them from the observed median temperatures to get residuals for the warm dust and hot gas temperatures. We show the residuals for the warm dust and hot gas temperatures on Figure~\ref{fig:TdustvskT}. We assign each point a color corresponding to the distance from R136. The residuals show no significant correlation between the hot gas and warm dust temperatures, suggesting that the quantities are not causally related. We discuss the implications further in Section~\ref{sec:Dust Mixing}.

\subsection{Comparison to H-alpha}
\label{subsec:Halpha_comp}

We compare the diffuse X-ray emission to the diffuse H$\alpha$ observations described in Section~\ref{sec:H-alpha Imaging} to explore turbulent mixing, evaporative conduction, and physical leakage at the hot-cold interface as stellar feedback loss mechanisms. 

\begin{figure*}
    \centering    \includegraphics[width=0.9\textwidth]{SB_R136.pdf}
    \caption{\textit{Image:} Composite image of diffuse H$\alpha$ in green and broad X-rays in blue. The annuli centered on R136 are shown in white. SB profiles were calculated using 60$^\circ$ segments of the annuli. We ignore the inner 5\arcsec\ from our analysis. R136 is marked with a cyan X. The image is 3.2\arcmin$\times$3.2\arcmin. North is up, and East is left. \textit{Plots a--f:} Normalized surface brightness as a function of radius from R136. Radius increases from 10\arcsec--70\arcsec\ (2.4--17~pc) by a factor of 5\arcsec\ (1.2~pc). These profiles highlight how the X-ray and H$\alpha$ emission vary across the shells and cavities surrounding R136 and identify where the hot gas is confined. Half of the profiles (b, c, and d) show X-ray peaks interior to the H$\alpha$ peaks, indicating partial confinement of the hot gas within the H$\alpha$ shell.}
    \label{fig:SB_R136}
\end{figure*}

\begin{figure*}
    \centering
\includegraphics[width=\textwidth]{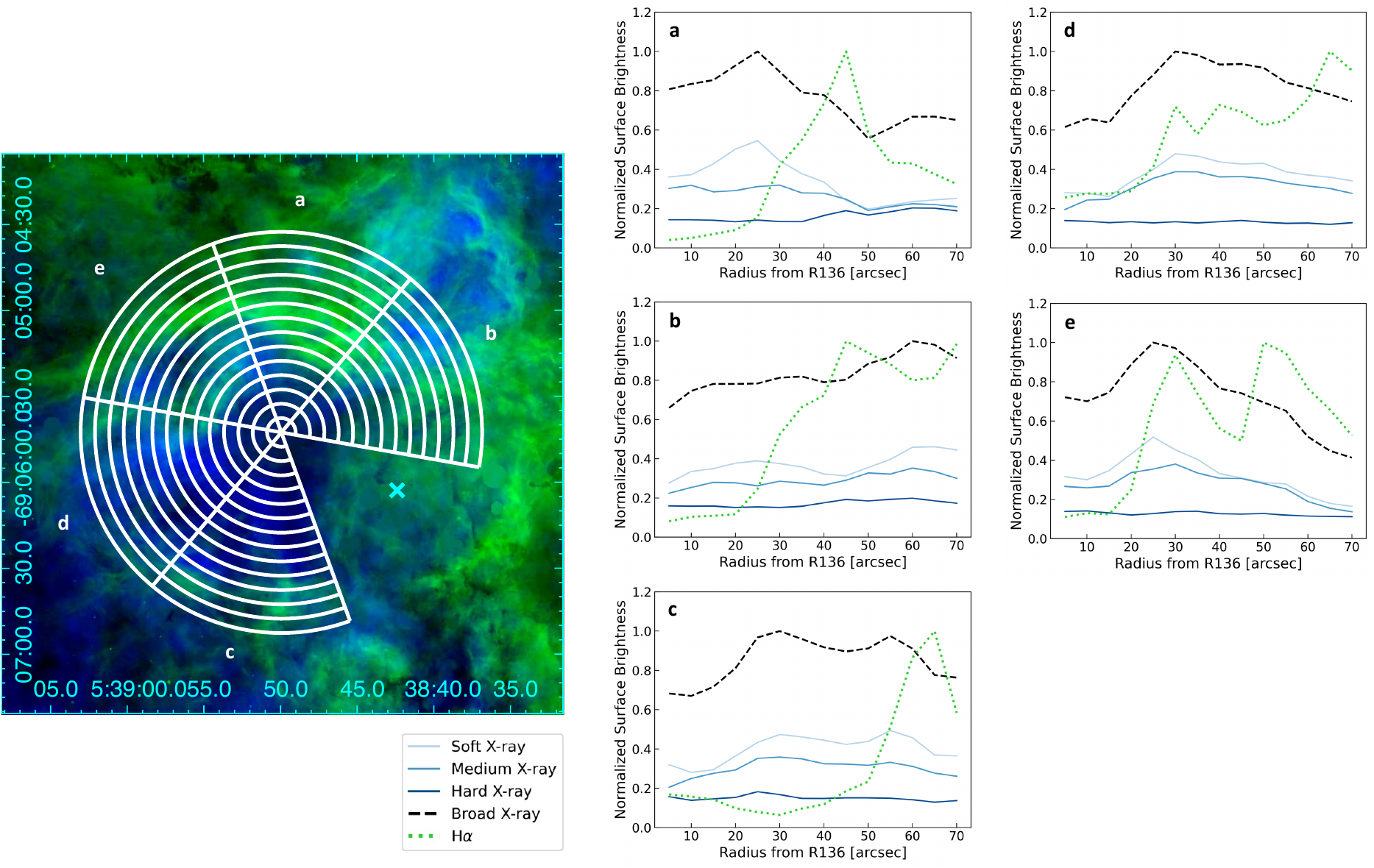}
    \caption{\textit{Image:} Composite image of diffuse H$\alpha$ in green and broad X-rays in blue. The annuli centered on the large cavity seen in the H$\alpha$ East of R136 are overplotted in white. SB profiles were calculated using 60$^\circ$ segments of the annuli (labeled \textit{a--e}, corresponding to the plots on the right). The section containing R136 was excluded from this analysis. The cyan X marks the location of R136. The image is 3.2\arcmin$\times$3.2\arcmin. North is up, and East is left. \textit{Plots a--e:} Normalized surface brightness as a function of radius from the center of the cavity. Radius increases from 5\arcsec--70\arcsec\ (1.2--17~pc) by a factor of 5\arcsec\ (1.2~pc). These profiles reveal the confinement of the X-rays surrounding the larger cavity by the varying H$\alpha$ shell. The X-rays peak interior to the H$\alpha$ peak in all profiles, indicating partial confinement of the X-rays. This is still the case for profiles with two H$\alpha$ peaks from neighboring cavities.}
    \label{fig:SB_largecavity}
\end{figure*}

\begin{figure*}
    \centering
\includegraphics[width=\textwidth]{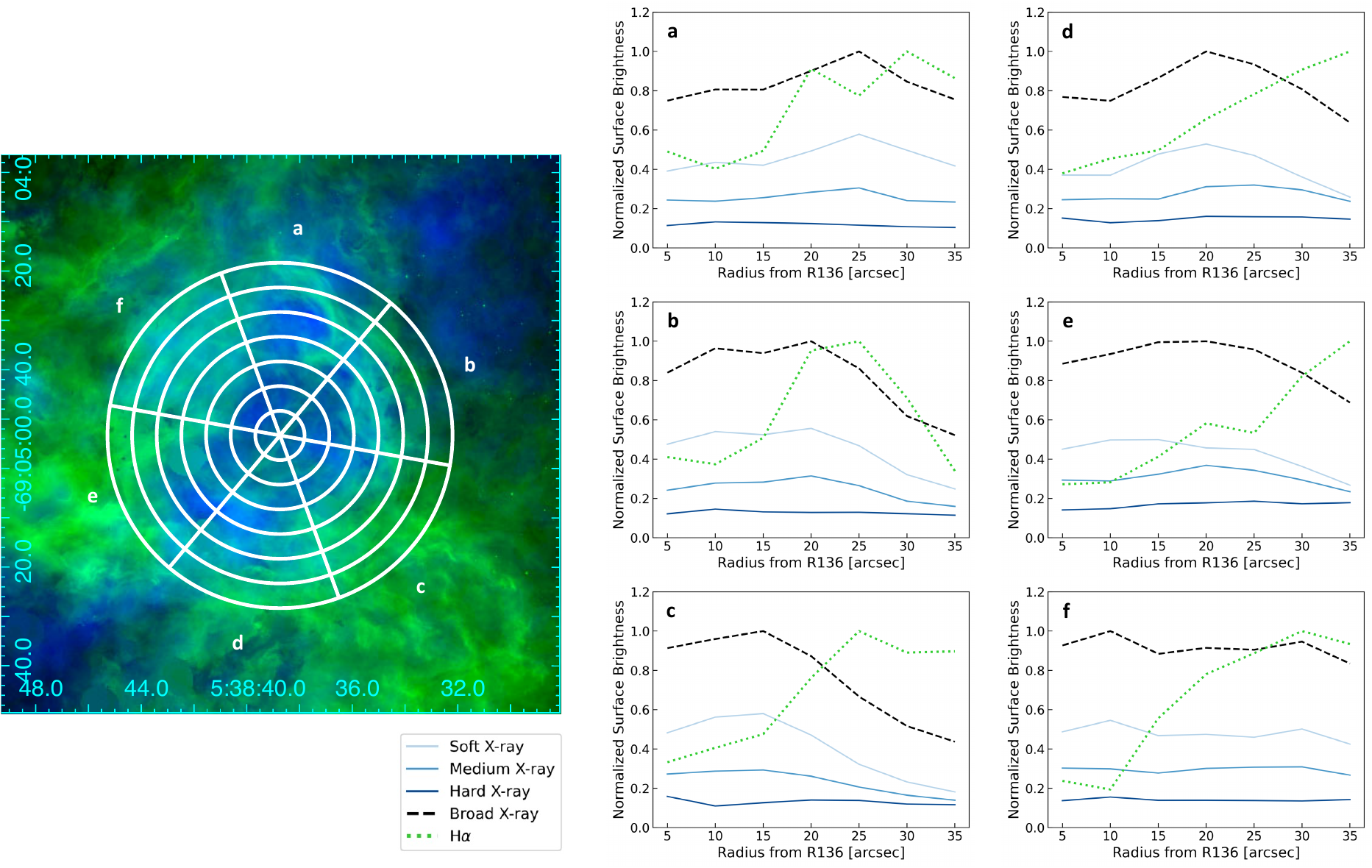}
    \caption{\textit{Image:} Composite image of diffuse H$\alpha$ in green and broad X-rays in blue. The annuli centered on the small cavity North of R136 are shown in white. SB profiles were calculated using 60$^\circ$ segments of the annuli. The image is 1.9\arcmin$\times$1.9\arcmin. North is up, and East is left. \textit{Plots a--f:} Normalized SB as a function of radius from R136. Radius increases from 5\arcsec--35\arcsec\ (1.2--8.5~pc) by a factor of 5\arcsec\ (1.2~pc). These profiles demonstrate the confinement of the X-rays surrounding the smaller cavity by the H$\alpha$ shell. In all profiles, the X-rays peak interior to the H$\alpha$ peak, suggesting partial confinement of X-rays.}
    \label{fig:SB_smallcavity}
\end{figure*}

We measure the soft X-ray (0.5$-$1.2~keV), medium X-ray (1.2$-$2~keV), hard X-ray (2$-$7~keV), broad X-ray (0.5$-$7~keV), and H$\alpha$ surface brightness (SB) profiles from three annuli centered on the magenta crosses in Figure~\ref{fig:hst_chandra}. The profiles are centered on R136 (Figure~\ref{fig:SB_R136}), the cavity east of R136 (Figure~\ref{fig:SB_largecavity}), and the cavity north of R136 (Figure~\ref{fig:SB_smallcavity}) to investigate the spatial relationship between the hot and warm gas. We calculate the profiles for R136 and the eastern cavity with annuli radii ranging from 5\arcsec--70\arcsec, while we calculate the northern cavity (which has a smaller radius) SB profiles with annuli of 5\arcsec--35\arcsec extents. We segment all annuli by 60$^{\circ}$ to explore the trends from X-ray and H$\alpha$ in different directions. Profiles depend on the segments we chose and may differ if other angles or segments had been selected. We normalize the H$\alpha$ and broad X-ray profiles using the maximum brightness of each profile for each of the segmented annuli. Soft, medium, and hard X-rays are scaled relative to the total broad-band X-ray profiles, such that the sum at each radius matches the normalized broad-band profile. In summary, almost all profiles show the X-rays peak interior to the H$\alpha$ in all three cavities, and the results are described in detail below.

The six SB profiles centered on R136 ($a-f$) are shown in Figure~\ref{fig:SB_R136}, excluding the inner 0\arcsec--5\arcsec\ due to unresolved point sources in R136 in the H$\alpha$ image. Segment $a$ overlaps with the cavity found in the H$\alpha$ to the north of R136. The broad X-rays peak at the location of the H$\alpha$ peak coincident with the H$\alpha$ arc before decreasing as the profile captures the cavity to the north of R136. Segment $b$ shows the most overlap between the H$\alpha$ and X-ray emission. The broad X-rays peak within 20\arcsec\ from R136 and are interior to the H$\alpha$ peak at 45\arcsec. Segment $c$ also shows overlap between the H$\alpha$ and X-rays, and a cavity is seen in the H$\alpha$ west of R136. The broad X-rays peak within 10\arcsec\ from R136, interior to the H$\alpha$ peak at 40\arcsec. Segment $d$ shows the broad X-rays peak 5\arcsec\ before the peak of the H$\alpha$ at 60\arcsec\ from R136. Segment $e$ overlaps with the cavity to the east of R136; the H$\alpha$ peaks at 10\arcsec\ from R136, and all X-ray profiles peak exterior to that. Segment $f$ overlaps with the cavity to the east and with the H$\alpha$ arc. The medium and hard X-rays remain nearly constant, while soft and broad X-rays increase with distance from the cluster. Overall for the segments where broad X-rays peak interior to H$\alpha$, the broad X-rays peak 5\arcsec--30\arcsec\ interior to the H$\alpha$ peak and they decrease by 5--40\% from the X-ray peak to the H$\alpha$ peak.

The five normalized SB profiles centered on the cavity located east of R136 ($a-e$) are shown in Figure~\ref{fig:SB_largecavity}. We exclude the segment overlapping with R136 from our analysis due to unresolved point sources surrounding the cluster in the H$\alpha$ image. All profiles show X-rays peaking interior to the H$\alpha$ peak. For most segments, the broad X-rays peak 10\arcsec--35\arcsec\ interior to the H$\alpha$ peak. Segment $b$ overlaps with the smaller cavity to the north of R136, hence the appearance of two H$\alpha$ peaks. The broad X-rays increase by 20\% before its peak within the H$\alpha$ peaks, suggesting the X-rays are not entirely confined. Segment $e$ also has two H$\alpha$ peaks as the profiles overlap with the H$\alpha$ arc and a second cavity. For all segments, the broad X-rays decrease by 20--30\% from their peak to the peak of the H$\alpha$.

The six normalized SB profiles centered on the smaller cavity located north of R136 ($a-f$) are shown in Figure~\ref{fig:SB_smallcavity}. The broad X-ray profiles peak interior to the H$\alpha$ peak in each segment. Profile $f$ remains mostly constant possibly due to it having the most overlap between H$\alpha$ and X-ray emission. The broad X-ray profiles peak 5\arcsec--20\arcsec\ interior to the H$\alpha$ peak, and they decrease by 5-35\% from the broad X-ray peaks to the H$\alpha$ peak.

Assuming the SB profiles represent sections of a sphere, we estimate a covering fraction for each segment to quantify the fraction of X-rays enclosed within the H$\alpha$ shells. For each profile, we integrate the broad X-ray SB interior to the radius at which the H$\alpha$ peaks, which we set as the location of the swept-up shell, and we compare this to the integral over the full radial extent to calculate the total X-ray SB. The resulting ratio provides an estimate of apparent confinement, though it should be interpreted with caution since projection effects and complex shell geometry may impact these values. For the profiles at R136, the northern cavity, and eastern cavity, the average confinement is 46\%, 55\%, and 73\%, respectively.

\begin{figure*}
    \centering   \includegraphics[width=\textwidth]{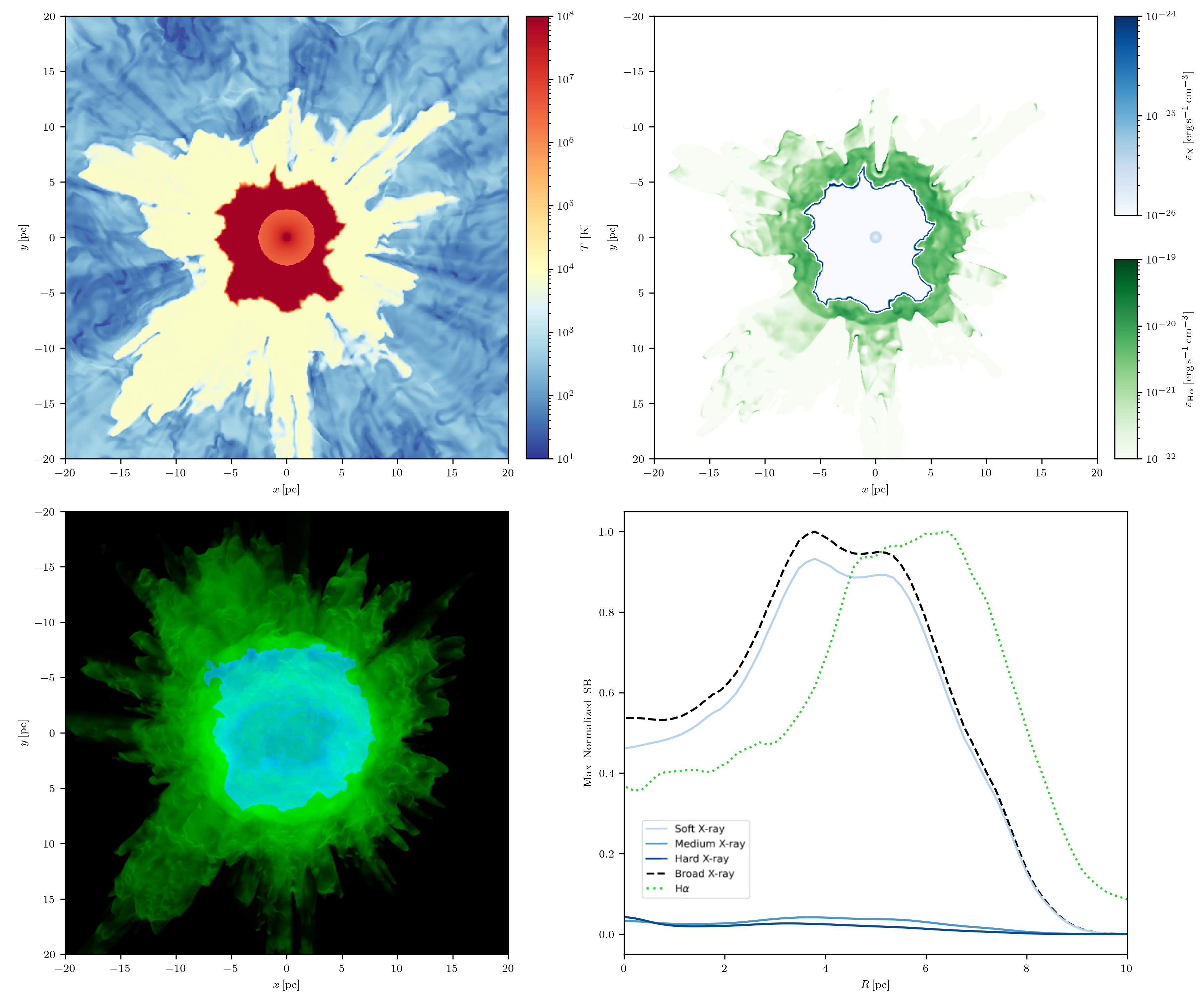}
    \caption{A mock observation of a star cluster of mass $3\times10^{3}~M_{\sun}$ feedback simulation presented in \citet{Lancaster25b}. \textit{Top Left}: A slice through the center of the 3D simulation of the gas temperature exhibiting the hot ($T\gtrsim 10^7\,{\rm K}$), shocked wind in red and the surrounding photo-ionized gas ($T\sim 10^4\, {\rm K}$) in yellow. \textit{Top Right}: The luminosity density of X-rays $\varepsilon_{\rm X}$ in blue (0.5$-$7 keV; top colorbar) and of H$\alpha$ $\varepsilon_{\rm H\alpha}$ in green (bottom colorbar). The X-ray emission is localized to near the WBB's surface. \textit{Bottom Left}: A two-color image of the X-ray surface brightness (blue) and H$\alpha$ surface brightness (green) in the simulation, both on a logarithmic scale. \textit{Bottom Right}: Analogous to the panels of Figures~\ref{fig:SB_R136}-\ref{fig:SB_smallcavity}, we show the X-ray and H$\alpha$ surface brightness profiles normalized to their maximum value as a function of radius. The simulations produce a central dip in the X-ray surface brightness and generate X-rays predominantly in the soft bands, in contrast to the relatively flat interior profiles with more comparable contributions across the soft, medium, and hard X-ray bands in Figures~\ref{fig:SB_R136}--\ref{fig:SB_smallcavity}. }
    \label{fig:simulation_plot}
\end{figure*}

\subsection{Simulation Comparison}
\label{subsec:sim_comp}

As a point of comparison and to guide the interpretation of our results above, we present a mock observation from simulations presented in \citet{Lancaster25b}. In Figure~\ref{fig:simulation_plot}, we show a variety of properties of the \texttt{MWR\_N512} simulation of \citet{Lancaster25b} of a $5\times 10^3\, M_{\odot}$ star cluster with a Kroupa IMF at solar metallicity. The simulation models the expansion of a feedback bubble driven by stellar winds and photo-ionizing radiation, which is treated as a point source at the center of the domain, with wind luminosities, wind mass-loss rates, and hydrogen-ionizing photon rates as calculated with \texttt{Starburst99} \citep{Leitherer92,SB99}. The bubble expands into a magnetized, turbulent medium with mean hydrogen number density $\overline{n}_{\rm H} = 100\,{\rm cm}^{-3}$. This simulation includes non-equilibrium hydrogen photo-chemistry of \citet{JGK_photchem}, allowing it to track the abundances of free-electrons and ionized hydrogen.

We use \texttt{CHIANTI} \citep{chianti97,chianti21} to calculate the X-ray emissivity as a function of wavelength and temperature, assuming the solar abundances of \citet{Scott15a,Scott15b}. The emissivity is integrated over the wavelengths corresponding to the soft ($0.5-1.2$~keV), medium ($1.2-2.0$~keV), and hard ($2.0-7.0$~keV) X-ray bands. This provides $j_{\rm X}/n_{\rm e} n_{\rm p}$, the X-ray emissivity ($j_{\rm X}$) per unit electron number density ($n_{\rm e}$) per unit proton number density ($n_{\rm }p$). Using the electron and proton densities, along with the hot gas temperature $kT$, we calculate the X-ray luminosity density shown in the upper right panel of Figure~\ref{fig:simulation_plot}. The fact that the simulation and the emissivity curves are calculated for solar metallicity are expected to effect the total X-ray luminosity, but not the spatial distribution of the X-rays in the simulation, which is the only thing we compare against below.

To simulate the H$\alpha$ emission, we use the effective rate coefficients for recombinations that result in H$\alpha$ emission as a function of temperature, $\alpha_{{\rm eff, H}\alpha}(T)$ (Equation 14.8 in \citealt{Draine_book}). The H$\alpha$ luminosity density is then:
\begin{equation}
    \Lambda_{{\rm H}\alpha}(n_e, n_p, T) = E_{{\rm H}\alpha}\alpha_{{\rm eff, H}\alpha}(T) n_e n_p
\end{equation}
where $E_{{\rm H}\alpha} = 1.89\, {\rm eV}$ is the energy of an H$\alpha$ photon. This is used to calculate the H$\alpha$ luminosity using $n_{\rm e}$, $n_{\rm p}$, and $kT$ as tracked in the simulation.

We calculate the X-ray and H$\alpha$ surface brightnesses by integrating the luminosity along the line of sight and dividing by the area represented by each pixel in the resulting map. We convolve the resulting X-ray surface brightness map with a two-dimensional Gaussian filter with standard deviations of 0.36~pc in order to approximately match the resolution of the derived X-ray image. The simulation resolution is larger (worse) than the resolution of the H$\alpha$ observation, so we perform no convolution for the H$\alpha$ mock observation. The surface brightness profiles shown in the bottom right panel of Figure~\ref{fig:simulation_plot} are constructed by averaging the maps in radial annuli and normalizing them to their peak values, following our analysis in Section~\ref{subsec:Halpha_comp}.

\section{Discussion} \label{sec:Discussion}

\subsection{Implications Regarding Dust Mixing} \label{sec:Dust Mixing}

As mentioned in Section~\ref{sec:intro}, turbulence at the hot-cold interface mixes hot gas with cooler ambient gas \citep{Breitschwerdt88, Fielding20}. With dust being mixed into the cold gas, turbulence may influence its distribution within the hot-cold interface and potentially its entrainment into the hot gas. The dust is heated by stellar radiation and, if embedded within the hot gas, by collisions and heat transfer with the hot gas, with sputtering and destruction of the dust grains occurring as a result \citep{Draine79, Rosen14}.

Both the hot gas and warm dust temperatures decrease with distance from R136 (Figure~\ref{fig:tempvsr}, yet neither follows the expected theoretical scaling relation of $T_{\rm dust} \propto r^{-2}$ for radiative heating or $kT \propto r^{-4/3}$ for the adiabatic expansion of a wind bubble. Both show a more gradual decline with distance from R136 than expected, which requires a more complex explanation.

The much more gradual decline in the warm dust temperature (compared to the hot gas) with distance from R136 could arise because the dust is not volume-filling and is confined to a thinner shell \citep{Draine11_dustyhii}. In this case, the dust emission projected closer to R136 may originate from dust at similar three-dimensional distances in other parts of the shell, resulting in similar dust temperatures throughout. The decline in X-ray temperature with distance from R136 suggests that a large fraction of the volume of the cavity is filled with a nearly isothermal, post-shock wind. This is consistent with our assumption of a volume filling factor of 1 (see Section~\ref{sec:spectralresults}).  The lack of correlation between the warm dust and hot gas temperatures in Figure~\ref{fig:TdustvskT} further indicates that dust heating is not dominated by collisional heating from the hot gas.

The significant scatter in the dust-mass-weighted radiation field intensity and the calculated dust temperature from \cite{Chastenet19} in Figure~\ref{fig:dust_properties} reflects local variations in the ISM structure, along with contributions from different heating sources or variations in intensity of a single heating source. The fraction of dust mass exposed to stronger radiation fields ($\gamma$) overall declines with distance from R136. However, there is scatter closer to R136 and $\gamma$ asymptotes at 150\arcsec\, suggesting that dust in the shell near R136 experiences a range of radiation intensities driven by the variations in the local ISM properties. Overall, our results are consistent with stellar radiation being the primary determinant of the dust temperature within 30 Dor.

\subsubsection{Caveats on Dust Mixing} \label{sec:caveats}

Figure~\ref{fig:maps} shows that the dust distribution from JWST partially overlaps with the regions of higher hot gas temperatures, which may indicate regions where the hot gas and dust could occupy the same volume. Given the shell geometry, the dust most likely lies in front of the hot gas or within the cooler swept-up material rather than being mixed into the hot gas. In these regions, we find a range of absorbing column densities, with higher $N_{\rm H}$ south of R136, which is consistent with dust along the line-of-sight rather than dust entrained in the hot gas. In the case that dust grains do mix into the hot phase, it may be subject to sputtering and destruction.

To assess whether dust grains could survive for extended periods in the hot gas before being sputtered, we calculate the approximate grain lifetime using Equation 25.14 in \cite{Draine_book}:
\begin{equation}
    \label{eq:dustlifetime}
    \tau_{\rm d} = 1\times10^5 \left[1+\left(\frac{T}{10^6\ {\rm K}}\right)^{-3}\right] \left(\frac{a}{0.1\ {\rm \mu m}}\right) \left(\frac{n_{\rm i}}{\rm cm^{-3}}\right)^{-1}~{\rm yr}.
\end{equation}

\noindent Using the X-ray temperatures and number densities from our spectral analysis (see Figure~\ref{fig:maps}) and assuming a typical grain size of $a = 0.1$~$\mu$m, we find that the dust grains immersed in the hot gas would have lifetimes of $\tau_{\rm d} \sim 0.5 - 2.25$~Myr and an average lifetime of $\tau_{\rm d} \sim0.85$~Myr. In a cluster like R136 (with an age of $\sim1-2$~Myr; \citealt{Hunter95, Crowther16, Bestenlehner20}), larger dust grains could survive and be entrained into the hot gas. However, the dust grains may not be significantly heated as the long timescales reflect the low collision rates in the hot gas, which drive both sputtering and energy transfer to the dust grains \citep{Draine79}. Since this timescale is dependent on the dust grain size and hot gas temperatures, a more detailed analysis of the dust grain distribution would be needed to assess the impact of sputtering and mixing into the hot gas on dust survival.

\subsection{Implications Regarding Turbulent Mixing} \label{sec:turbulent mixing}

Cooling is efficient at the interface layer between the shock-heated gas and the swept-up cooler ISM. Dynamical instabilities at the hot-cold interface drive turbulence and subsequent mixing, allowing for efficient radiative cooling to occur in the intermediate temperatures ($\sim10^5$~K) where the radiative cooling function peaks \citep{Draine_book}. Thermal conduction and evaporation drive energy into the cooler shell and cooler gas into the bubble interior respectively \citep{ElBadry19, Lancaster21a, Lancaster21b}. The analytic model from \cite{Weaver77} includes thermal conduction and cooling losses, which allows for evaporation of the shell into the hot bubble interior. However, due to the assumption of a smooth, spherical shell, it does not capture the turbulent mixing and instability-driven mass exchange seen in more recent simulations. Updated models, such as \cite{ElBadry19}, implement conduction and mixing processes, leading to weaker evaporation rates and enhanced cooling compared to the Weaver solution. 

The normalized H$\alpha$ SB profiles centered on R136 in Figure~\ref{fig:SB_R136} exhibit clear limb brightening, where the surface brightness increases toward the shell. In contrast, the X-rays do not show limb brightening. Instead, the broad X-ray profiles remain relatively flat, peaking interior to the H$\alpha$, and decline as the H$\alpha$ increases. This behavior can be explained by cooling near the hot-cold interface and is consistent with expectations from evaporative conduction. Alternatively, this trend in the profiles could occur due to a narrow region of shocked wind material which resides close to the outer shell. This would indicate substantial energy losses.

We can see from the top right panel of Figure~\ref{fig:simulation_plot} that the X-ray emission in these simulations is dominated by the mixing layer at the interface between the hot wind-blown bubble (WBB) and the surrounding photo-ionized gas. This fact leads to the X-ray surface brightness profile (the bottom right panel of Figure~\ref{fig:simulation_plot}) to be ``limb brightened'' with a clear cavity in the interior. This trend is not seen as clearly in the observations, which show relatively flat surface brightness profiles interior to the outer peaks. The simulations also differ from the observations in that nearly all the X-ray SB is provided by the soft band, whereas the observations have relatively equal contribution from each band (though decreasing somewhat at higher energy). 

Both of these discrepancies could be explained by the lack of thermal conduction in the simulations of \citet{Lancaster25b}. Conduction would lead to mass-loading of the hot, shocked gas interior to the bubble, leading to more emission in the interior (reducing the limb brightening seen in the simulations) and more emission at higher temperatures (leading to more contribution from the harder X-ray bands). The problem of lack of production of hard X-rays could also be alleviated by inverse Compton scattering of the local radiation field by cosmic-ray electrons, as has been suggested recently in the context of the circum-galactic medium \citep{Hopkins25_ICXraysa} and cool-core clusters \citep{Hopkins25_ICXraysb}. However, predictions for star-forming regions in this context have not yet been carried out.

\subsection{Other Loss Channels: Leakage} \label{sec:leakage}

Due to the fractal density structure of the surrounding ISM, hot wind gas may escape through low-density channels, leading to partial leakage from the bubble shell \citep{HC09,Lancaster21c}. The comparison of the diffuse X-ray and H$\alpha$ SB profiles indicates the varying levels of X-ray confinement by the H$\alpha$ shell (see Figures~\ref{fig:SB_R136} $-$ \ref{fig:SB_smallcavity}). For all three profiles, we find that most of the broad X-rays peak interior to the H$\alpha$ peaks and that the former decrease as the H$\alpha$ increases. Around R136, half of the broad X-ray profiles in Figure~\ref{fig:SB_R136} (segments b, c, and d) peak 5\arcsec\ $-$ 30\arcsec\ ($\approx 1 - 7$~pc) interior to the H$\alpha$ peak, with the normalized broad X-ray SB decreasing by $5-40$\% from their maxima to the H$\alpha$ peak. 

The segments with X-ray peaks interior to the H$\alpha$ peaks suggest stronger confinement, while those where X-rays peak at or beyond the H$\alpha$ peaks demonstrate uneven expansion and gaps in the shell from where the X-rays may be leaking from the \hii\ region. This is supported by the non-negligible confinement fractions calculated in Section~\ref{subsec:Halpha_comp}, which average 46\%$-$73\% across the profiles.

The structure of the shells and cavities around R136 also suggest the presence of hot gas leakage. The H$\alpha$ arcs and cavities are not uniformly spherical and instead exhibit fragmented morphologies as the hot gas expands. This creates channels through which the hot gas can escape, while at the same time driving turbulent mixing where the X-ray and H$\alpha$ emission align. This is also evident in the simulations in Figure~\ref{fig:simulation_plot}, which show the complex, non-uniform distribution of the H$\alpha$ and X-ray emission and indicate the possibility that leakage and turbulent mixing arise from the interaction of the hot gas and the surrounding ISM.

\section{Conclusions} \label{sec:conclusions}

In this paper, we analyze the $\sim2$~Ms archival Chandra data of 30 Doradus to explore stellar wind energy loss channels. We extract and model X-ray spectra of 84 regions of 42\arcsec$\times$42\arcsec, corresponding to the Spitzer SEDs (Figure~\ref{fig:84regions}) for comparison to dust properties, and we find the best-fit hot gas properties ($kT$, $n_{\rm X}$, and $N_{\rm H}$) from fitting of the X-ray spectra in the $0.5-3$~keV range (Figure~\ref{fig:maps}). We extract additional spectra from 60 regions (Figure~\ref{fig:hst_chandra}) to explore the hot gas properties in relation to the H$\alpha$ cavities and shells around R136.

We compare the X-ray and dust temperatures with distance from the R136 (Figures~\ref{fig:tempvsr} and ~\ref{fig:dust_properties}) and distance-corrected residuals (Figure~\ref{fig:TdustvskT}). While both temperatures are higher near R136 and decline with distance, their radial trends arise from different processes. The distance-corrected residuals show no significant correlation, which suggests that the dust is likely confined to the shell around the wind-blown bubble and primarily heated by the strong stellar radiation field from R136.

We also compare normalized X-ray and H$\alpha$ surface brightness profiles from segmented annuli centered on R136 and two cavities seen in the H$\alpha$ to the north and east of R136 (Figures~\ref{fig:SB_R136} -\ref{fig:SB_smallcavity}). We find most profiles have X-ray emission peaking interior to the H$\alpha$ shell, and the X-rays declines toward the edge of the shells. This indicates that the hot gas is partially confined with confinement fractions of 46\%$-$73\%. 

Additionally, we compare our profiles to recent simulations of wind blown bubbles including turbulent mixing (Figure~\ref{fig:simulation_plot}). These simulations find X-ray emission peaks interior to the H$\alpha$ shells, but the simulations have enhanced X-ray emission at the shell edge and insufficient emission in the cavity interior compared to our observations. These differences may arise from the lack of thermal conduction in the simulations, which would lead to mass-loading of (and therefore enhanced emission from) the interior of the hot wind bubble. While confirming this suggestion will require comparison against models that include conduction, this comparison clearly illustrates the power of the X-ray and H$\alpha$ observations as diagnostics of the complex physics of massive star-forming regions.

\begin{acknowledgements}
The authors would like to thank Matthew Povich, Adam Leroy, and Tony Wong for helpful discussions related to the work. J.A.R., L.A.L., and S.L. acknowledge support through the Heising-Simons Foundation grant 2022-3533. L.A.L. and L.L. also gratefully acknowledge the support of the Simons Foundation, with L.L. supported under grant 965367. 
\end{acknowledgements}

\software{CIAO (v4.14; \citealt{CIAO2006}), XSPEC (v12.11.1; \citealt{XSPEC})}

\bibliographystyle{aasjournal}
\bibliography{bibliography}{}

\end{document}